\documentclass[twocolumn]{aastex61}
\usepackage{hyperref}

\shorttitle{\lowercase{$r$}-process and kilonovae}

\shortauthors{Wanajo}

\begin{document}

\title{Physical conditions for the \lowercase{$r$}-process I. radioactive energy sources of kilonovae}

\newcommand{\X}[1]{ #1}

\author{Shinya Wanajo}
\affil{Department of Engineering and Applied Sciences,
Sophia University, Chiyoda-ku, Tokyo 102-8554, Japan;
shinya.wanajo@sophia.ac.jp}
\affil{iTHEMS Research Group, RIKEN, Wako, Saitama 351-0198, Japan}
\affil{Max Planck Institute for Gravitational Physics (Albert Einstein Institute), Am M\"uhlenberg 1, Potsdam-Golm, D-14476, Germany}

\begin{abstract}
Radioactive energies from unstable nuclei made in the ejecta of neutron star mergers play principal roles in powering kilonovae. In previous studies power-law-type heating rates (e.g., $\propto t^{-1.3}$) have frequently been used, which may be inadequate if the ejecta are dominated by nuclei other than the $A\sim 130$ region. We consider, therefore, two reference abundance distributions that match the $r$-process residuals to the solar abundances for $A \ge 69$ (light trans-iron plus $r$-process elements) and $A \ge 90$ ($r$-process elements). Nucleosynthetic abundances are obtained by using free-expansion models with three parameters: expansion velocity, entropy, and electron fraction. Radioactive energies are calculated as an ensemble of weighted free-expansion models that reproduce the reference abundance patterns. The results are compared with the bolometric luminosity ($>$ a few days since merger) of the kilonova associated with GW170817. We find that the former case (fitted for $A \ge 69$) with an ejecta mass $0.06\, M_\odot$ reproduces the light curve remarkably well including its steepening at $\gtrsim 7$~days, in which the mass of $r$-process elements is $\approx 0.01\, M_\odot$. Two $\beta$-decay chains are identified: $^{66}$Ni$\rightarrow^{66}$Cu$\rightarrow^{66}$Zn and $^{72}$Zn$\rightarrow^{72}$Ga$\rightarrow^{72}$Ge with similar halflives of parent isotopes ($\approx 2$~days), which leads to an exponential-like evolution of heating rates during 1--15~days. The light curve at late times ($> 40$~days) is consistent with additional contributions from the spontaneous fission of $^{254}$Cf and a few Fm isotopes. If this is the case, the event GW170817 is best explained by the production of both light trans-iron and $r$-process elements that originate from dynamical ejecta and subsequent disk outflows from the neutron star merger.
\end{abstract}


\section{Introduction}
\label{sec:introduction}

The discovery of a binary neutron star (NS) merger as the source of gravitational wave signals confirmed by advanced LIGO/Virgo \citep[on August 17, 2017; GW170817,][]{Abbott+2017}, followed by the detection of an electromagnetic counterpart across the entire wavelength range, has opened a window of the ``multi-messenger astronomy" \citep[see, e.g.,][]{Metzger2017}. Among the latter, the observations of electromagnetic emission in optical and near-infrared ranges \citep[kilonova,][]{Li1998, Metzger2010} provided us with numerous clues to understanding the origin of heavy elements such as gold and uranium made by the rapid neutron-capture process \citep[$r$-process, see][for a recent review]{Thielemann2017}.

The history of the $r$-process study goes back to the seminal works by \citet{Burbidge1957} and \citet{Cameron1957}. \citet{Burbidge1957} supposed a sort of stellar explosions including core-collapse supernovae and Type~Ia supernovae be the astrophysical site of the $r$-process. Prior to these works, in \citet{Burbidge1956} a concept of the $r$-process already has been presented by associating the light curves of Type~Ia supernovae decaying in an exponential manner with the spontaneous fission of a transuranic species $^{254}$Cf (with a halflife of 60.5 days). It was before the identification of the decay chain $^{56}$Ni$\rightarrow^{56}$Co$\rightarrow^{56}$Fe \citep{Colgate1969, Arnett1979} that actually powers Type~Ia supernovae.

Since then, a main focus of the study of $r$-process sites has been placed on core-collapse supernovae including prompt explosions \citep{Schramm1973, Sato1974, Hillebrandt1976, Sumiyoshi2001, Wanajo2003} and more recently neutrino-driven explosions \citep{Meyer1992, Woosley1994, Witti1994, Qian1996, Cardall1997, Otsuki2000, Wanajo2001, Thompson2001}. However, recent studies exclude the former explosion mechanism \citep[e.g.,][]{Kitaura2006, Janka2012} and suggest the latter be sources of only light trans-iron elements \citep[e.g.,][]{Wanajo2011, Wanajo2013}. Currently, only a magneto-rotationally induced explosion mechanism remains a viable possibility among the scenarios of core-collapse supernovae \citep{Winteler2012, Nishimura2015, Nishimura2017, Moesta2017, Halevi2018}. 

An alternative scenario, the decomposition of neutron-rich material from merging binary neutron stars (NSs, or a NS and a black hole) also was proposed as the sources of $r$-process elements \citep{Lattimer1974, Symbalisty1982, Eichler1989, Meyer1989}. Early studies of the dynamical ejecta (resulting from tidal torque and shock heating) of NS mergers pointed the robust production of heavy $r$-process nuclei ($A > 120$) by a fission recycling in extremely neutron-rich environments \citep{Freiburghaus1999, Goriely2011, Korobkin2012, Bauswein2013}. More recent studies indicate, however, the production of all the $r$-process species ($A \ge 90$) in the ejecta with a wide range of neutron-richness owing to weak interactions \citep[electron/positron and neutrino captures on free nucleons,][]{Wanajo2014, Sekiguchi2015, Sekiguchi2016, Goriely2015, Radice2016, Papenfort2018}. The post-merger outflows from the accretion disk orbiting around a remnant (a massive NS or a black hole) also were suggested as a site of the $r$-process \citep{Ruffert1999, Metzger2008, Surman2008, Wanajo2012}. Recent studies indicate various possibilities such that the disk outflows are modestly \citep[e.g.,][]{Just2015, Fujibayashi2018} or very \citep[e.g.,][]{Wu2016, Siegel2017, Fernandez2018} neutron-rich. 

Latest studies of Galactic chemical evolution imply that $r$-process-enhanced halo stars reflect the nucleosynthetic yields from single NS merger events \citep{Ishimaru2015, Beniamini2016, Safarzadeh2017, Ojima2018}. Spectroscopic analyses of Galactic halo stars also have been providing us with several important clues. The remarkable agreement of the abundance distributions of $r$-process-enhanced halo stars with those of the solar $r$-process component indicates a robust and single origin of $r$-process elements (e.g., CS~22092-052, \citealt{Sneden2003}; CS~31082-001, \citealt{Siqueira2013}). The stars observed in a recently discovered ultra-faint dwarf galaxy, Reticulum~II, also exhibit solar-like $r$-process abundance patterns \citep{Ji2016, Ji2018}. However, such a remarkable agreement appears not to extend to the low-$Z$ ($< 50$) and high-$Z$ ($> 80$) ends; abundance variations of a factor of several can be seen in lighter elements \citep[e.g., Sr, Y, and Zr,][]{Sneden2008} as well as in actinides \citep[e.g., Th,][]{Holmbeck2018} with respect to those in between (e.g., Eu). More seriously, there are few measurable lines of light trans-iron elements ($30 < Z < 40$) that compose the low-mass side of the $r$-process residuals to the solar system abundances \citep[e.g.,][]{Goriely1999}. Currently, only Ga and Ge have been measured from the space for only a few halo stars \citep{Sneden2008}.

The identification of a kilonova (AT~2017gfo or SSS17a) associated with GW170817 has revealed the production of elements beyond iron in the NS merger \citep[e.g.,][]{Arcavi2017, Chornock2017, Cowperthwaite2017, Kasen2017, Kasliwal2017, Nicholl2017, Pian2017, Smartt2017, Tanaka2017, Tanvir2017}. While the early ``blue" emission indicates a lanthanide-free ($Z < 57$) component in the ejecta \citep{Metzger2014, Kasen2015, Tanaka2018}, the late-time ($>$ a few days) emission in red-optical and near-infrared wavelengths confirms the presence of freshly synthesized lanthanides that have high opacities  \citep{Banes2013, Kasen2013, Tanaka2013}. However, the inferred mass fraction of lanthanides and heavier in the ejecta is only $\approx 10^{-4}$--$10^{-2}$ \citep[e.g.,][]{Arcavi2017, Chornock2017, Nicholl2017, Waxman2017}. It is questionable, therefore, if the merger made heavy $r$-process elements such as gold and uranium. Moreover, such photometric analyses alone cannot discriminate between lanthanides and heavier elements. 

Another problem is the large amount of ejecta from the merger; the inferred masses of the blue and red components are, respectively, $\approx 0.01$--$0.02\, M_\odot$ with the outflow velocity of $\approx 0.2\, c$ ($c$ is the speed of light) and $\approx 0.04\, M_\odot$ with $\approx 0.1\, c$ \citep[e.g.,][]{Cowperthwaite2017, Nicholl2017}. The total mass $\approx 0.05$--$0.06\, M_\odot$ is too large to be fulfilled by the dynamical ejecta of $\lesssim 0.01\, M_\odot$ \citep[][]{Hotokezaka2013, Bauswein2013, Sekiguchi2016, Radice2016}. The disk outflows may eject more material of $\sim 0.01$--$0.1\, M_\odot$ but with smaller velocity \citep[$\sim 0.05\, c$,][]{Dessart2009, Metzger2014, Just2015, Siegel2017, Shibata2017, Fujibayashi2018}. Note
that most of the above estimates for the kilonova ejecta were based on the power-law-type heating rates \citep[e.g., $\approx 2\times 10^{10}\, t^{-1.3}$~erg~g$^{-1}$~s$^{-1}$, where $t$ is time in days,][]{Metzger2010, Wanajo2014} originating from the decaying radioactivities with $A\sim 130$.

In this paper we revisit the issue of the radioactive heating rates in NS merger ejecta, which is supposed to be the first of the series of papers that explore the physical conditions for the $r$-process by using a multi-component free-expansion model described in \S~\ref{sec:model}. Nucleosynthetic abundances are obtained by using free-expansion models that cover a wide range of parameters (expansion velocity, entropy, and electron fraction). The heating rates are then calculated as an ensemble of free-expansion models with their weighted abundances, which fit the $r$-process residuals to the solar system abundances \citep{Goriely1999} for two cases: a) $A\ge 69$ and b) $A\ge 90$ (\S~\ref{sec:reference}). Obviously, the choice of reference abundance distributions are not unique; these two cases are taken for simplicity, in which the nuclei of $A \sim 70$ and 130, respectively, play dominant roles for radioactive heating. The resultant heating rates are presented in \S~4 with discussion on the contributions from $\beta$-decay, $\alpha$-decay, and fission. In \S~5, the heating rates for the two cases by adopting the thermalization efficiencies in \citet{Barnes2016} are compared with the kilonova light curve of the NS merger GW170817. Summary and conclusions follow in \S~6.

\section{Multi-component free-expansion model}
\label{sec:model}

First, we define a free expansion (FE) model to be used throughout this study. Provided that a spherically symmetric, homogeneous gaseous matter adiabatically expands with time $t$, the temporal evolution of matter density is given by
\begin{equation}
    \label{eq:density}
    \rho(t) = \rho_0 \left(1 + \frac{t}{R_0/v}\right)^{-3},
\end{equation}
where $\rho_0 = 1.4\times 10^9$~g~cm$^{-3}$ and $R_0 = 150$~km are taken as the density and radius at $t = 0$ \citep[a similar approach can be seen in][]{Freiburghaus1999, Farouqi2010}. Although the FE model (hereafter FE) itself is site-independent, these boundaries are chosen according to the result of the hydrodynamical simulation of a NS merger in \citet{Wanajo2014}. The radial expansion velocity $v$ in Eq.~(\ref{eq:density}) is assumed to be constant, which is one of free parameters in a FE described below.

An ensemble of FEs constitutes a multi-component FE (mFE) model such that the nucleosynthetic abundances satisfy
\begin{equation}
    \label{eq:mfe}
    Y_i = \sum_{j=1}^{N_\mathrm{FE}} \phi_j Y_{\mathrm{FE}, i, j},
\end{equation}
where $Y_i$ is the abundance of the $i$th isotope in the mFE, $Y_{\mathrm{FE}, i, j}$ the abundance of the $i$th isotope in the $j$th FE, and $\phi_j$ a weight for the $j$th FE \citep[see a similar approach in][]{Bouquelle1996, Goriely1996}. A set of $\phi_j$'s will be determined in \S~\ref{sec:reference}.

Each FE involves three free parameters, namely, a constant expansion velocity $v$, an initial entropy $S$ (in units of Boltzmann constant per nucleon, $k_\mathrm{B}/\mathrm{nuc}$), and an initial electron fraction (proton-to-nucleon ratio) $Y_\mathrm{e}$. In this study the ranges of these parameters are taken to be $(v/c, S, Y_\mathrm{e}) = $ (0.05--0.30, 10--35, 0.01--0.50) with the intervals of $(\Delta(v/c), \Delta S, \Delta Y_\mathrm{e}) = $ (0.05, 5, 0.01). These cover the ranges in the bulk of dynamical ejecta \citep[e.g.,][]{Wanajo2014} and disk outflows \citep[e.g.,][]{Fujibayashi2018}. In Eq.~(\ref{eq:mfe}), therefore, the total number of FEs is $N_\mathrm{FE} = 6\times 6\times 50 = 1800$.

Nucleosynthetic abundances for each FE are obtained by using a nuclear reaction network code, {\tt rNET},  described in \citet{Wanajo2001, Wanajo2014}. {\tt rNET} consists of 6300 isotopes of $Z = 1$--110 with experimental rates when available (e.g., JINA REACLIB V2.0\footnote{https://groups.nscl.msu.edu/jina/reaclib/db/index.php.}, \citealt{Cyburt2010}; Nuclear Wallet Cards\footnote{http://www.nndc.bnl.gov/wallet/}) and theoretical estimates otherwise (e.g., TALYS, \citealt{Goriely2008} for neutron, proton, and $\alpha$ captures and GT2, \citealt{Tachibana1990} for $\beta$-decays with the HFB-21 mass prediction, \citealt{Goriely2010}). Neutrino captures are not included in this study. Theoretical (spontaneous, neutron-induced, and $\beta$-delayed) fission properties adopted are those predicted from the HFB-14 mass model \citep{Goriely2007}. A single Gaussian-type distribution of fission fragments is assumed with a prompt emission of four neutrons per event. For the energy released per fission, an empirical law of average total kinetic energies
\begin{equation}
    \label{eq:fission}
    \langle \mathrm{TKE} \rangle = 0.1189\times \frac{Z^2}{A^{1/3}} + 7.3\quad \mathrm{MeV}
\end{equation}
is taken from \citet{Viola1985}. For $1550 \le Z^2/A^{1/3} \le 1650$, in which measured energies are appreciably greater than those in Eq.~(\ref{eq:fission}) ($A \sim 260$ in the present case), $\langle \mathrm{TKE} \rangle = 250$~MeV is adopted \citep[see, e.g., Figure~20 in][]{Hessberger2017}.

The nucleosynthesis calculation for each FE starts when the temperature decreases to 10~GK. The initial composition is determined to be $1 - Y_\mathrm{e}$ and $Y_\mathrm{e}$ for free neutrons and free protons, respectively, which immediately attains nuclear statistical equilibrium (NSE) in such high temperature. The temperature in each timestep is computed from the density in Eq.~(\ref{eq:density}), the entropy, and the isotopic composition by using a tabulated equation of state \citep{Timmes2000}. The entropy generation from $\beta$-decay, $\alpha$-decay, and fission at each timestep is taken into account.

\section{Determination of abundance distributions}
\label{sec:reference}

\begin{deluxetable}{cccccccc}
\tabletypesize{\scriptsize}
\tablecaption{Properties of mFE Models}
\tablewidth{0pt}
\tablehead{
\colhead{Model} &
\colhead{Reference $A$\tablenotemark{a}} &
\colhead{$X_\mathrm{r}$\tablenotemark{b}} &
\colhead{$X_\mathrm{l}$\tablenotemark{c}} &
\colhead{$X_{66}$\tablenotemark{d}} &
\colhead{$X_{72}$\tablenotemark{e}} &
\colhead{Th/Eu\tablenotemark{f}} &
}
\startdata
mFE-a & 69--205 & 0.15 & 0.035 & 0.035 & 0.0073 & 0.84 \\ 
mFE-b & 88--205 & 0.72 & 0.21 & 0.0044 & 0.0014 & 0.83 \\ 
\enddata
\tablenotetext{a}{Range of the atomic mass number of the residuals adopted for the recurrence method in Eq.~(\ref{eq:recurrence}).}
\tablenotetext{b}{Mass fraction of $r$-process elements ($A \ge 90$).}
\tablenotetext{c}{Mass fraction of lanthanides and heavier ($A \ge 139$).}
\tablenotetext{d}{Mass fraction of $^{66}$Zn (the daughter of $^{66}$Ni).}
\tablenotetext{e}{Mass fraction of $^{72}$Ge (the daughter of $^{72}$Zn).}
\tablenotetext{f}{Production ratio in number.}
\label{tab:properties}
\end{deluxetable}

\begin{figure*}
\epsscale{1.17}
\plottwo{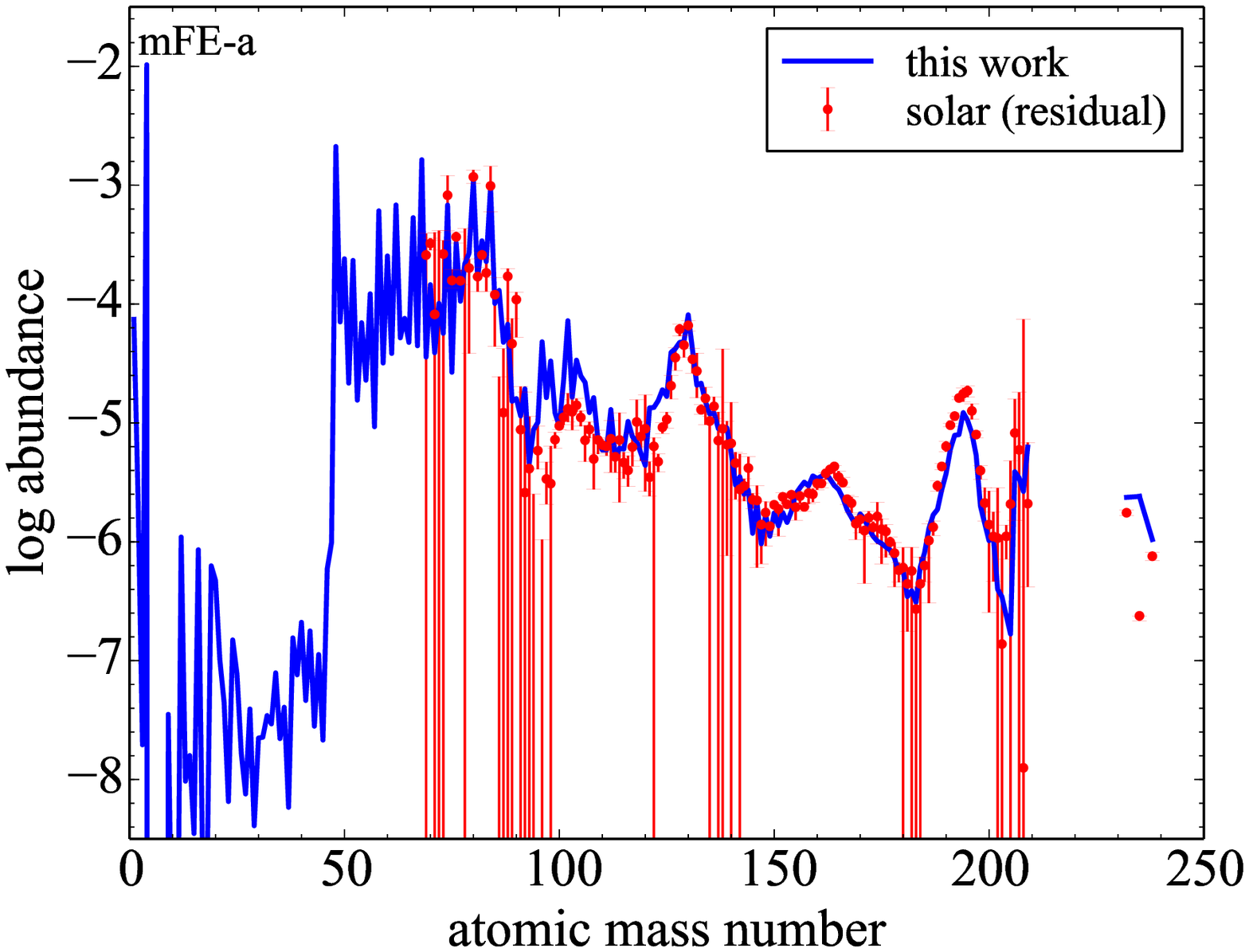}{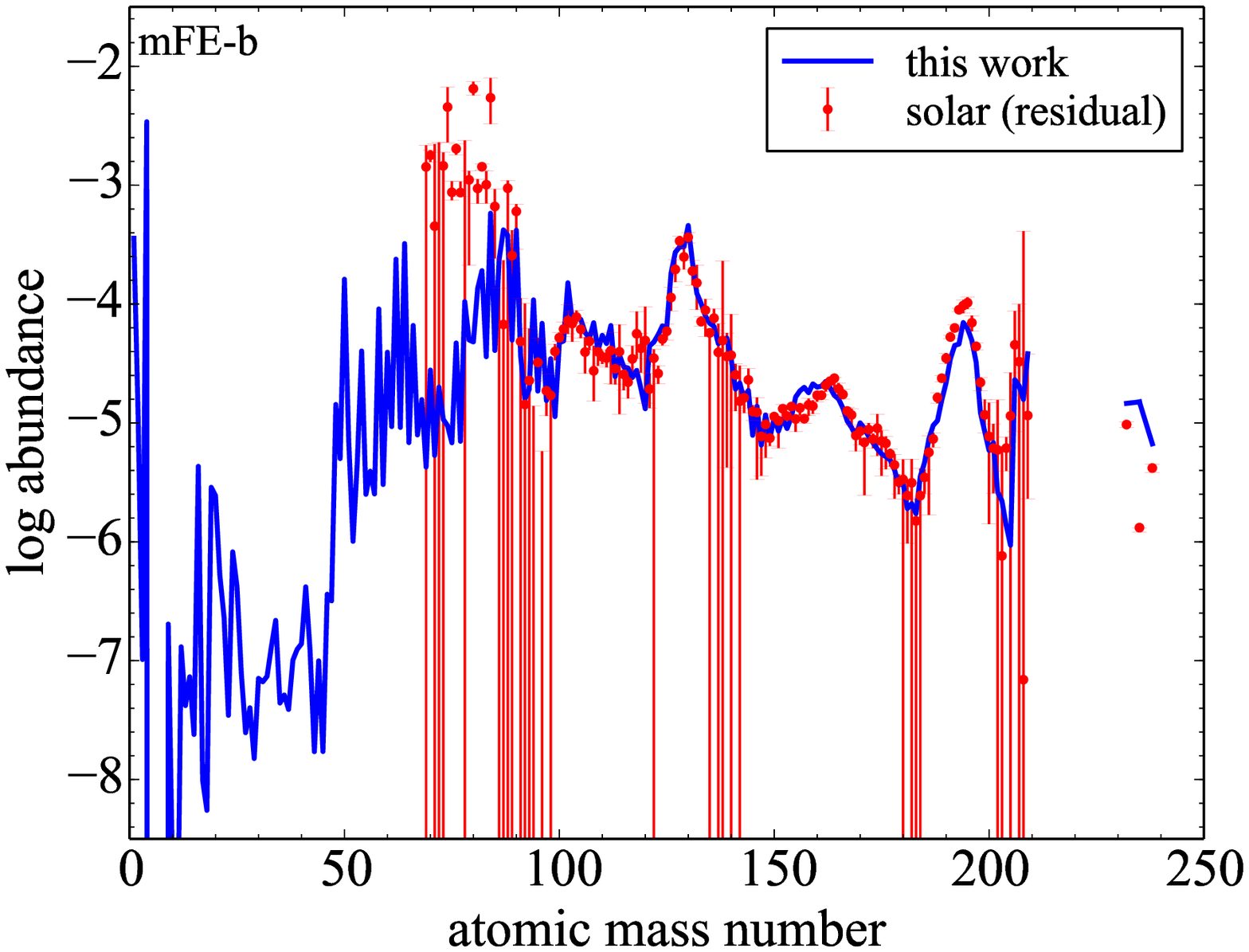}
\plottwo{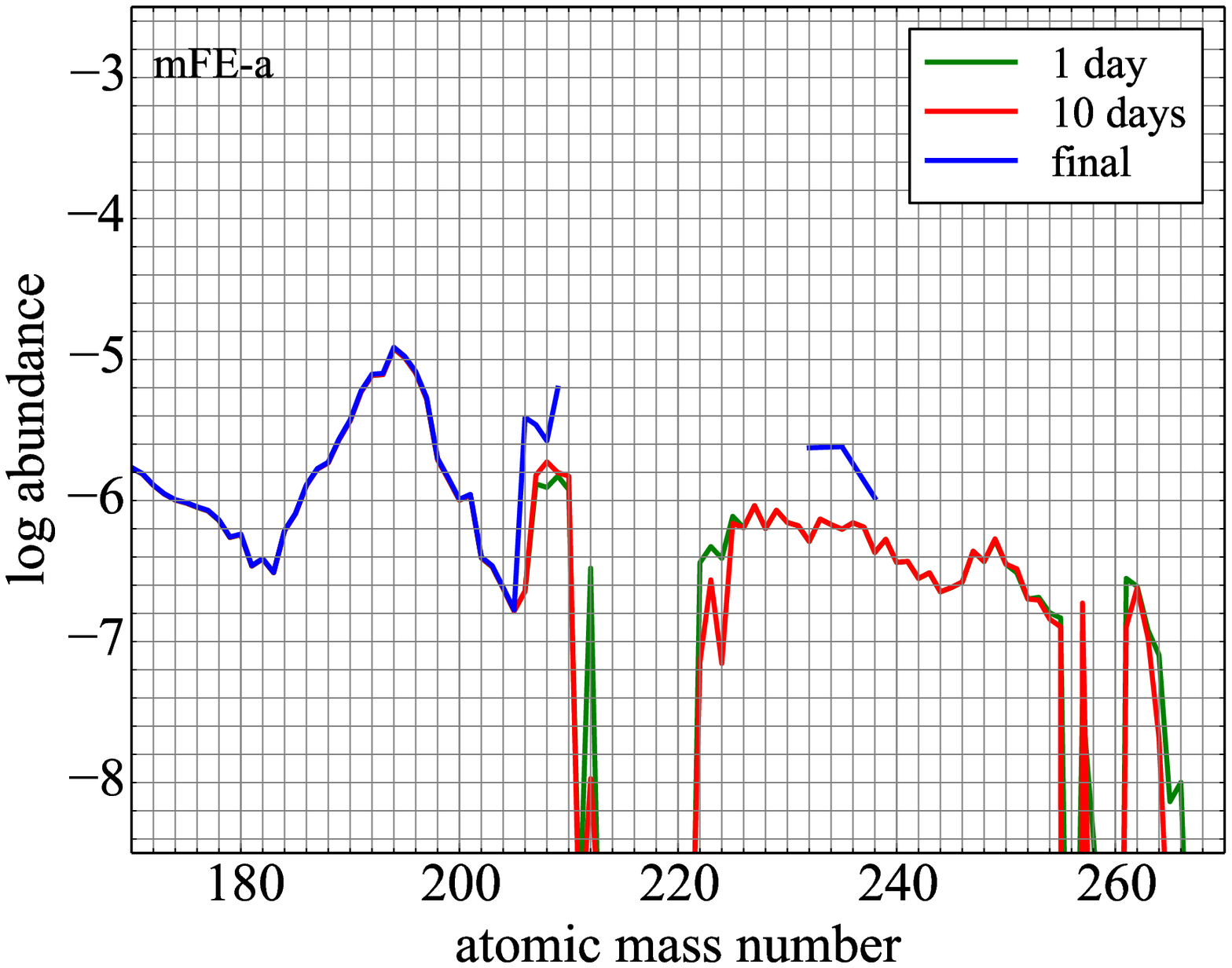}{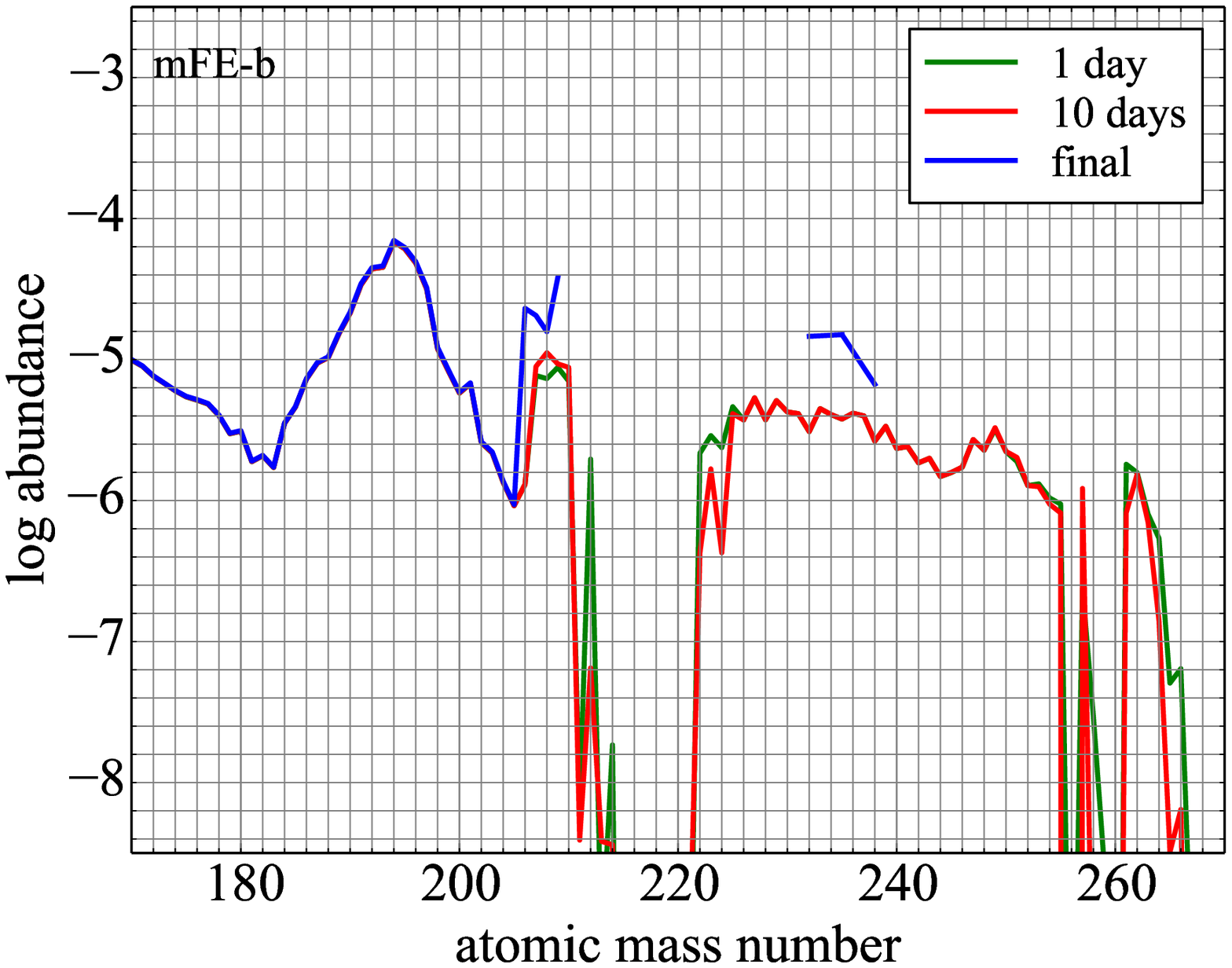}

\caption{\textit{Top}: Final abundances (blue lines) of mFE-a (left) and mFE-b (right) with the $r$-process residuals to the solar system abundances (red circles with error bars, \citet{Goriely1999}) as functions of atomic mass number. The residual abundances are shifted to match those of mFEs at $A = 138$. \textit{Bottom}: Abundances of mFE-a (left) and mFE-b (right) as functions of atomic mass number at 1 (green) and 10 (red) days along with the final abundances (blue).
}
\label{fig:nuclei}
\end{figure*}

\begin{figure*}
\epsscale{1.17}
\plottwo{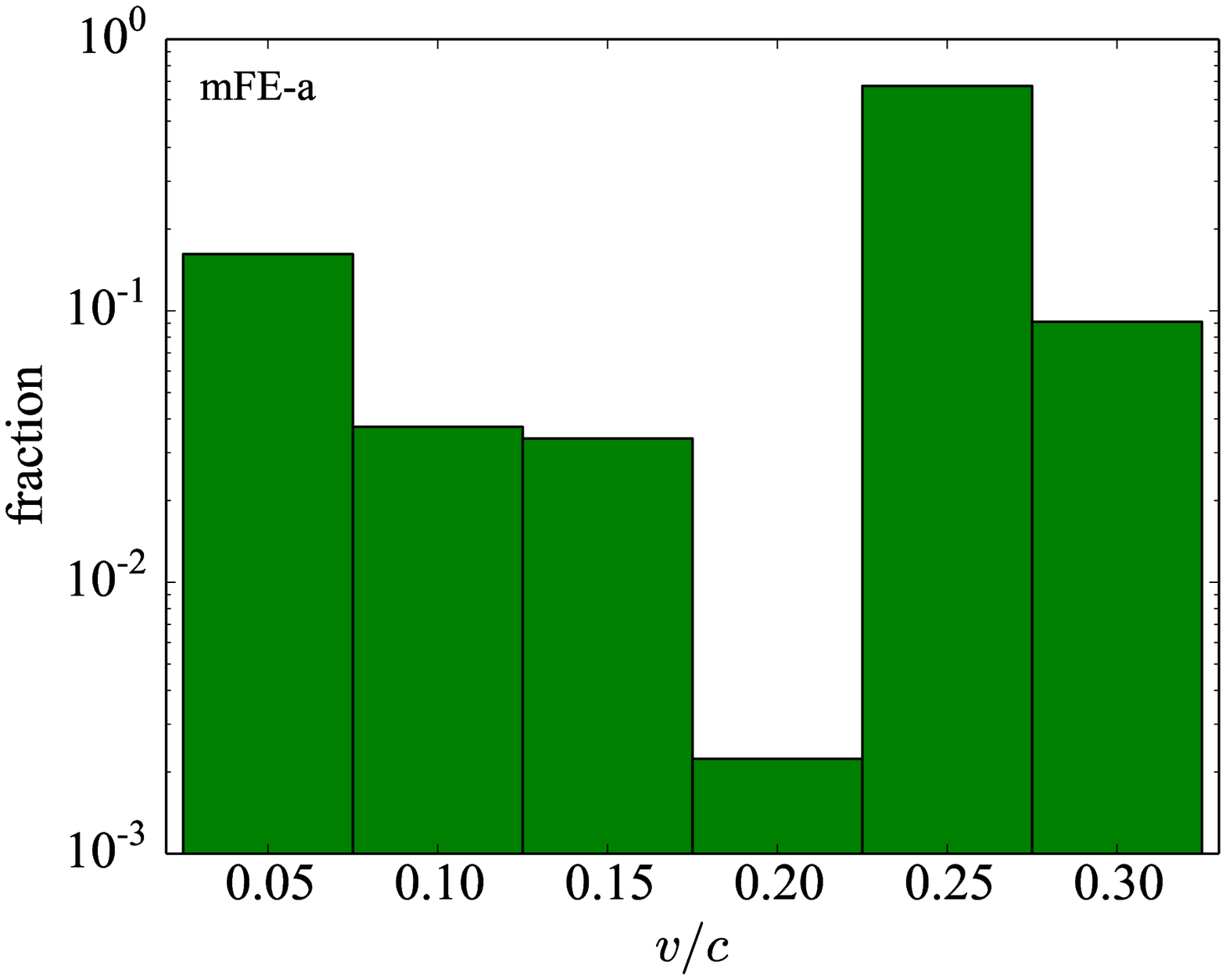}{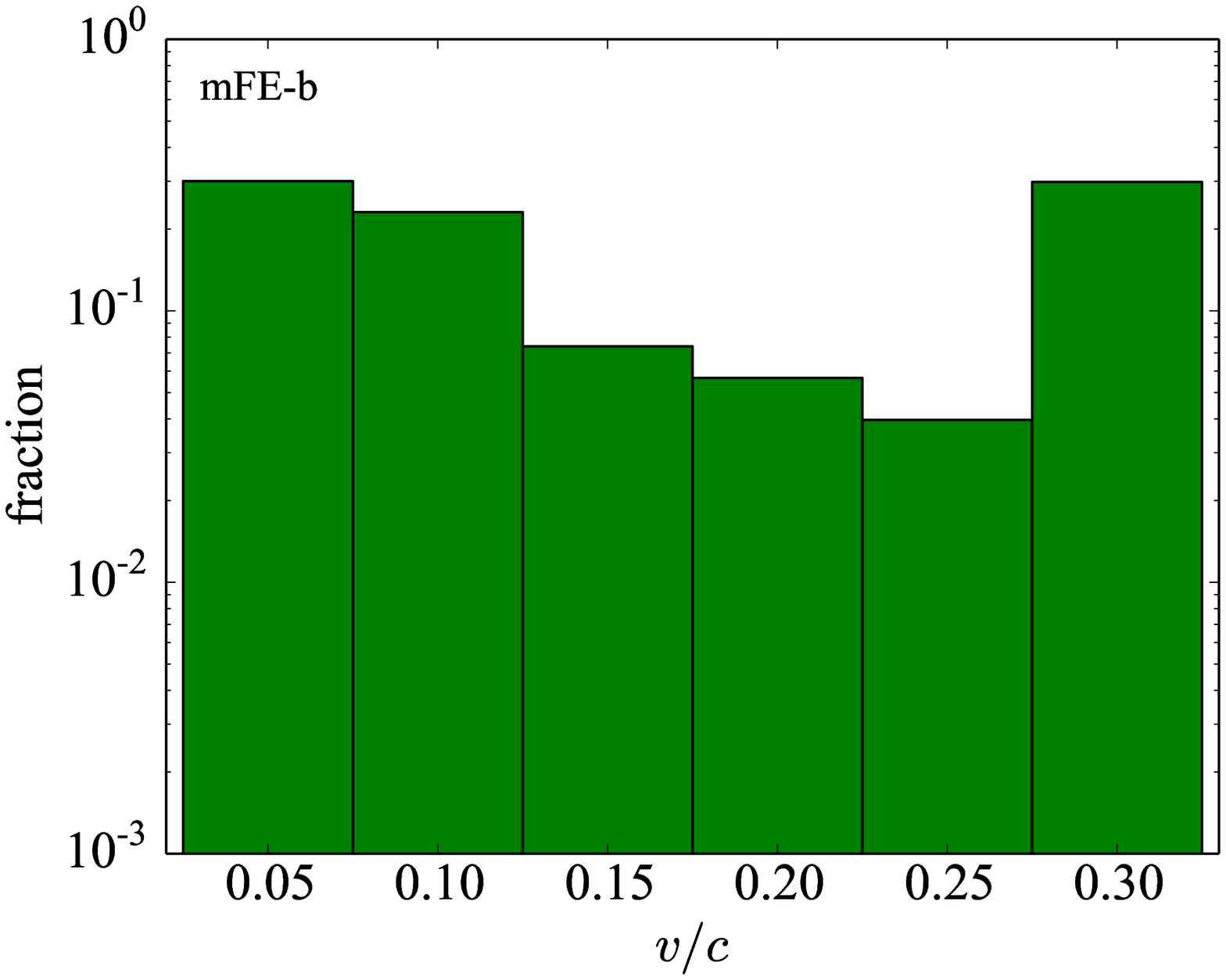}
\plottwo{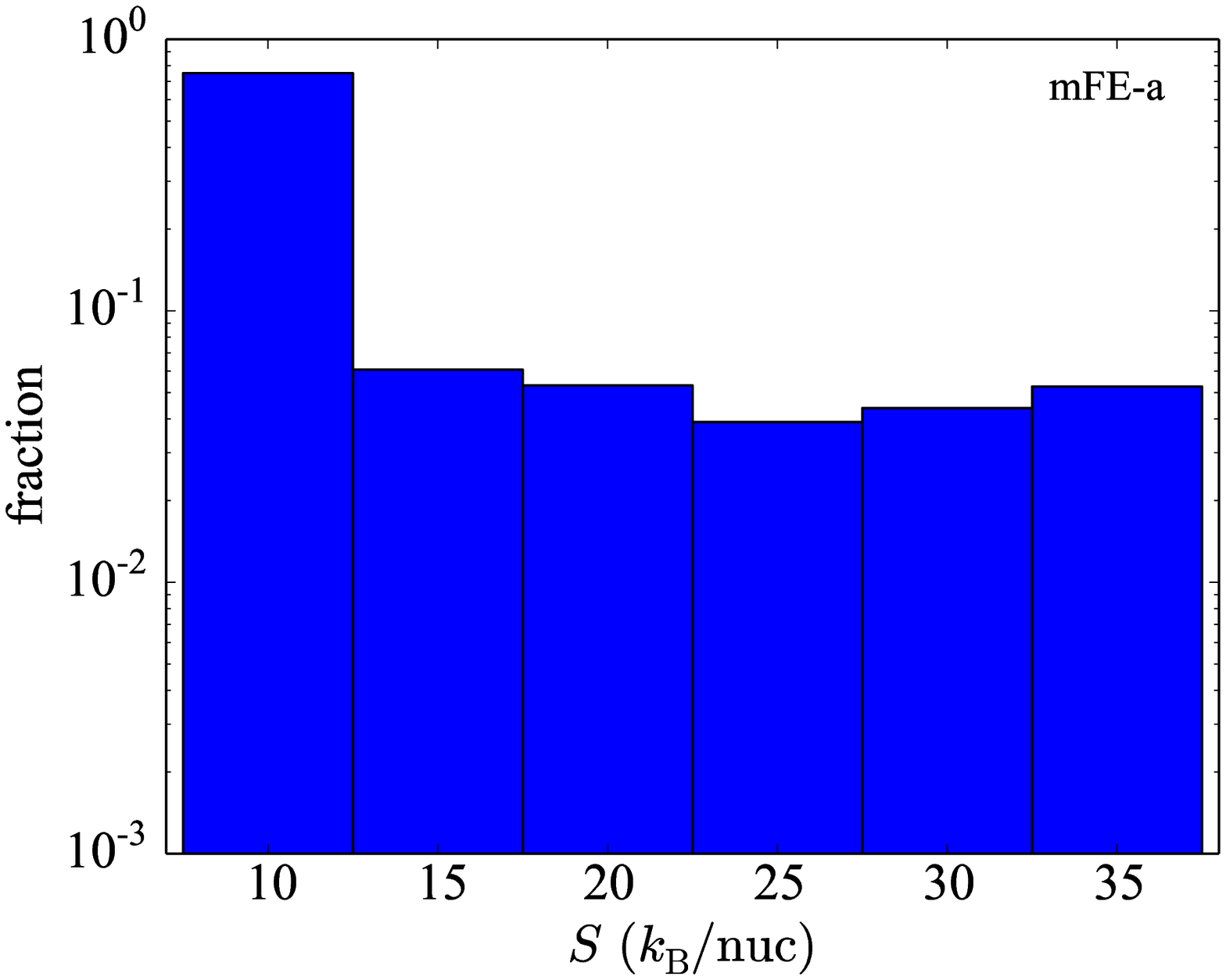}{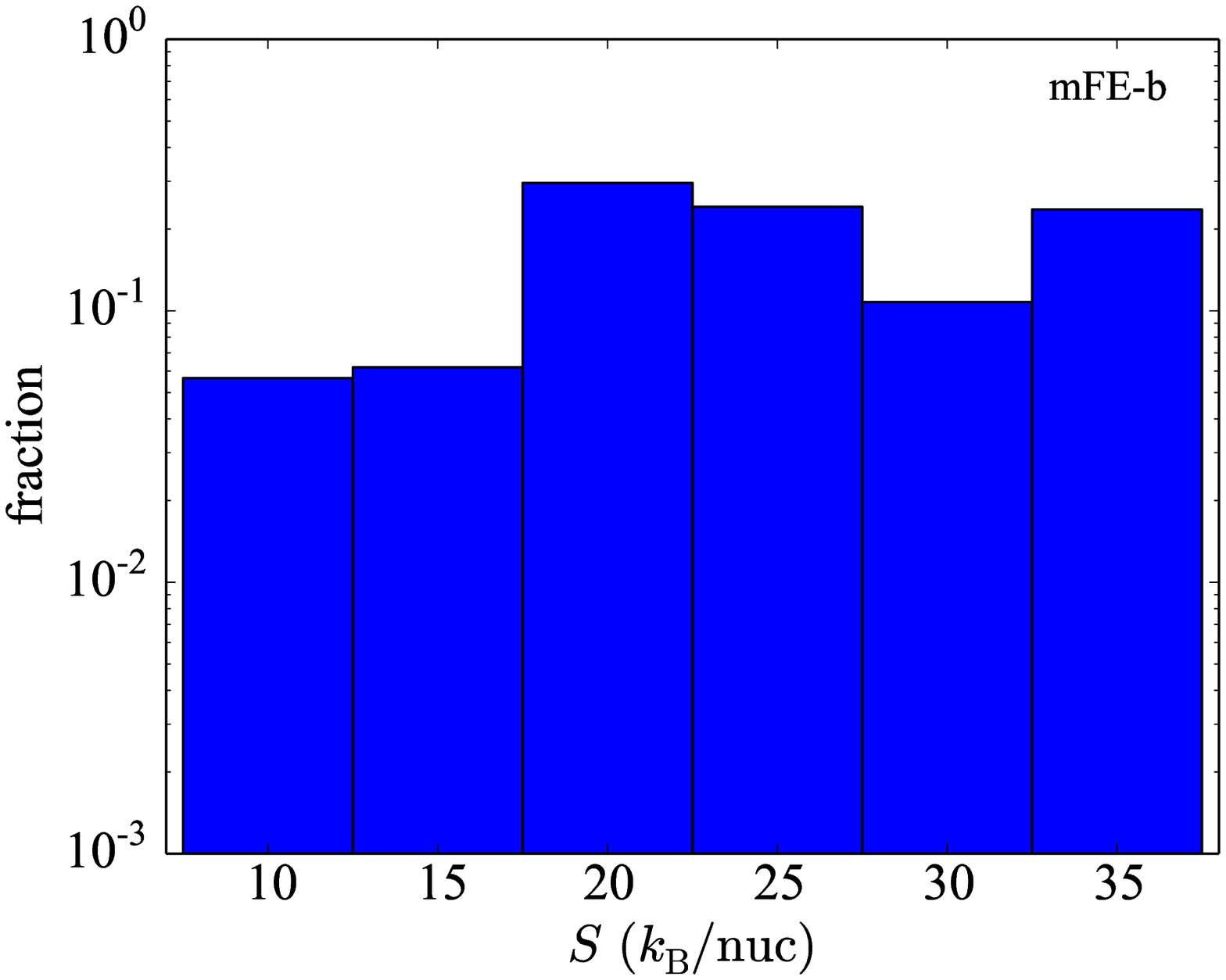}
\plottwo{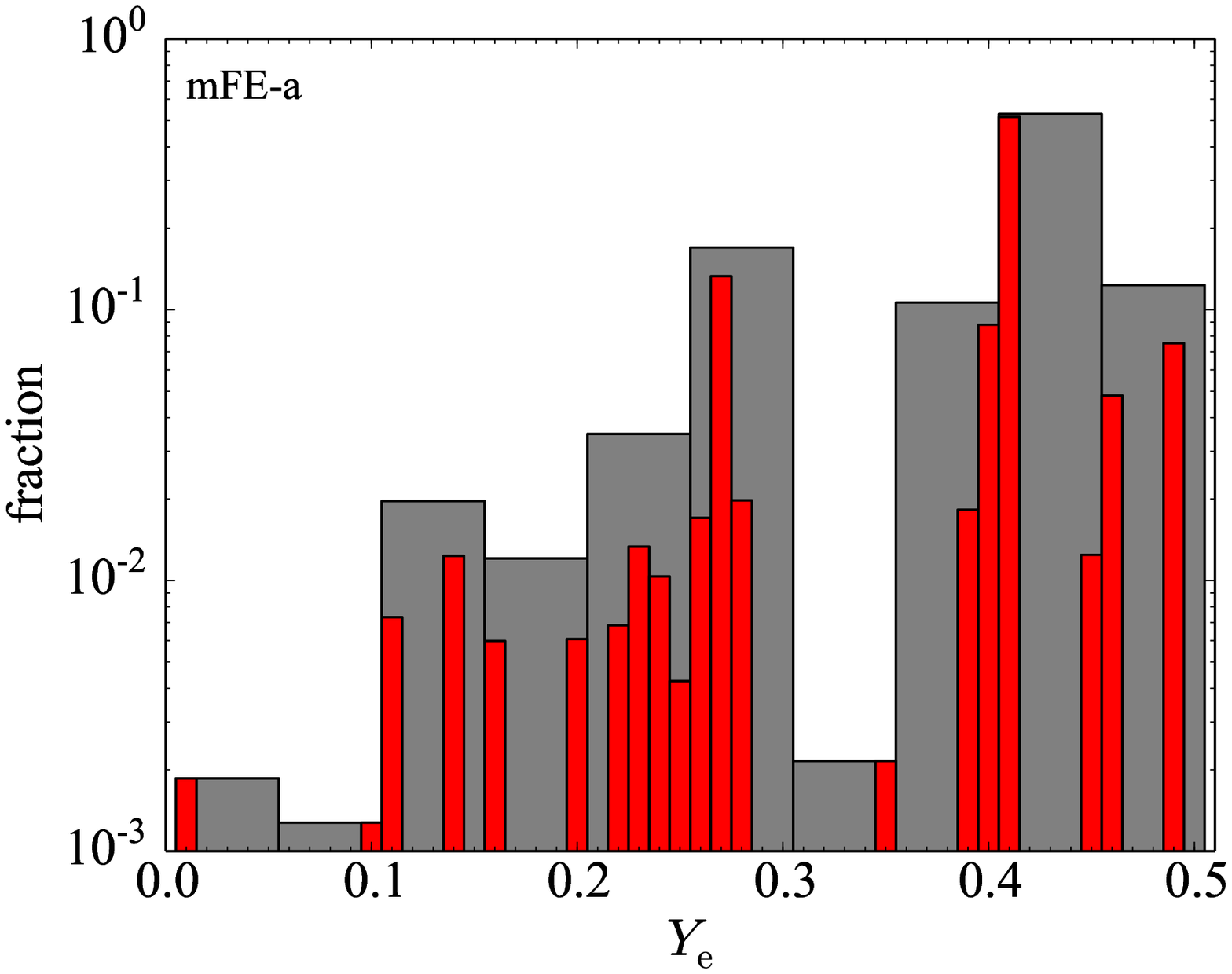}{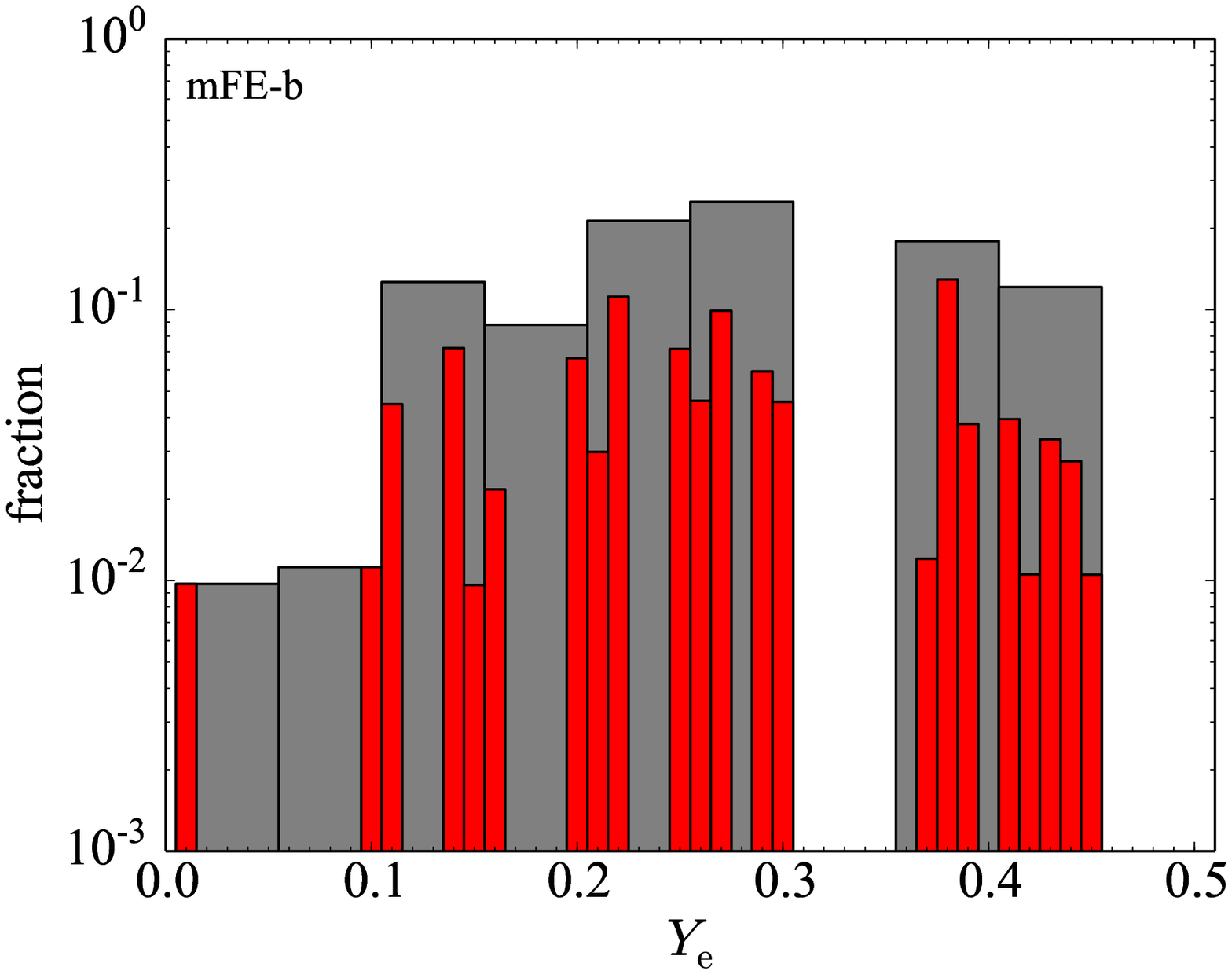}

\caption{Fractions of $v/c$ (top), $S$ in units of $k_\mathrm{B}/\mathrm{nuc}$ (middle), and $Y_\mathrm{e}$ (bottom) over the parameter ranges for mFE-a (left) and mFE-b (right). In the bottom panels, also shown in grey are the histograms grouped with an interval of $\Delta Y_\mathrm{e} = 0.05$. Note that the distributions of these quantities presented here should not be regarded as the unique solution (see text).
}
\label{fig:ye}
\end{figure*}

To obtain the weights $\phi_j$'s in Eq.~(\ref{eq:mfe}), a reference abundance distribution should be chosen. For this purpose, we employ the $r$-process residuals to the solar system abundances (by subtracting the $s$-process component) as a function of $A$ \citep[$= 69$--238, red circles with error bars in Figure~\ref{fig:nuclei},][]{Goriely1999}. This may be justified because of the robustness of the abundance distributions in $r$-process-enhanced stars, which match the solar $r$-process pattern at least in the range $50 < Z < 80$. Note that the ``$r$-process residuals" with $A < 90$ are merely those made by nucleosynthesis other than the $s$-process. In fact, it is known that such nuclei can be produced in NSE under low $S\, (\sim 10\,k_\mathrm{B}/\mathrm{nuc})$ and modestly low $Y_\mathrm{e}\, (\sim 0.4)$ conditions \citep{Hartmann1985, Meyer1998, Wanajo2018}. To clarify this point, we refer to the components of $A < 90$ and $A \ge 90$ as ``light trans-iron elements" and ``$r$-process elements", respectively, and an ensemble of both as ``residuals".

For the reference abundance distribution, mFE is constructed by using a recurrence method described in \citet{Bouquelle1996},
\begin{equation}
    \label{eq:recurrence}
    \phi_j^{k+1} = \phi_j^k\, \sum_i^{N_\mathrm{nuc}} \frac{Y_{\mathrm{ref}, i}}{Y_i^k}\, Y_{\mathrm{FE}, i, j}
\end{equation}
with the initial weights of $\phi_j^1 = 1/N_\mathrm{FE}$, where $Y_{\mathrm{ref}, i}$ is the $i$th abundance of the reference and $i$ runs over all reference nuclei with the total number $N_\mathrm{nuc}$. The abundances are normalized such as  $\sum Y_i^k = 1$, $\sum Y_{\mathrm{FE}, i, j} = 1$, and $\sum Y_{\mathrm{ref}, i} = 1$, where $k$ indicates the $k$th trial.  The iteration is terminated when $|\phi_j^{k+1}-\phi_j^k| < 10^{-4}$ is satisfied.

In light of little information for the abundances of light trans-iron elements in $r$-process-enhanced stars, we consider two mFEs, refer to as mFE-a and mFE-b hereafter, with the minimum a) $A = 69$ \citep[as in][]{Goriely1999} and b) $A = 88$, respectively (Table~\ref{tab:properties}, second column). The former represents the residual abundances of both light trans-iron and $r$-process elements and the latter $r$-process elements only. For both cases, those of $A > 206$ are not included for the fitting procedure, which do not represent the yields at a nucleosynthetic event (but at the formation of the solar system). The determined nuclear abundances of mFE-a and mFE-b are displayed in Figure~\ref{fig:nuclei} (upper left and upper right, respectively) with those of the residuals that are shifted to match at $A = 138$. For both cases, mFEs reasonably reproduce the residual distribution over the adopted $A$ range. Note the co-production of abundances that are out of range ($A < 69$ and $A < 88$ for mFE-a and mFE-b, respectively, and $A > 205$ for both). 

While the abundance patterns are similar to each other for $A \ge 90$, mFE-a results in 5 times smaller $r$-process products. In fact, the mass fraction of $r$-process elements ($A \ge 90$) in mFE-a is only $X_\mathrm{r} = 0.15$ (3rd column in Table~\ref{tab:properties}) because of the dominance of light trans-iron elements. As a result, the mass fraction of lanthanides and heavier ($A \ge 139$) is $X_\mathrm{l} = 0.035$ (4th column in Table~\ref{tab:properties}), which is comparable to the upper bound of estimates for GW170817 in the literature \citep[$\approx 0.01$,][]{Chornock2017}. In contrast, mFE-b gives large mass fractions of $r$-process elements ($X_\mathrm{r} = 0.72$) and of lanthanides and heavier ($X_\mathrm{l} = 0.21$); the latter is appreciably greater than the literture values.

The bottom panels of Figure~\ref{fig:nuclei} display the trans-lead abundances at 1 and 10 days for mFE-a (left) and mFE-b (right). The abundance patterns are similar to each other and thus the final Th/Eu ratios are almost the same (7th column in Table~\ref{tab:properties}). The abundances with $A > 266$ already have decayed away by spontaneous fission of parent nuclei in the neutron-rich region. Transuranic nuclei up to $A = 266$ remain including $^{254}$Cf and some Fm isotopes that contribute to the heating rates by spontaneous fission. Noted that these fissile nuclei are absent in \citet[][in their Figure~1]{Barnes2016}. The reason of discrepancy is likely due to the different fission barriers that determine the lifetimes by spontaneous fission \citep[see large variations of theoretical fission barriers in Figure~2 and resulting fissile regions in Figure~8 of][]{Goriely2015b}. 

Figure~\ref{fig:ye} shows the fractions of  $v/c$ (top), $S$ (middle), and $Y_\mathrm{e}$ (bottom) over the parameter ranges for mFE-a (left) and mFE-b (right). For $Y_\mathrm{e}$, such discrete distributions may be unphysical and due to numerical reasons (in which a small number of FEs are preferentially taken in the recurrence relation). For the purposes of subsequent discussion, the $Y_\mathrm{e}$ fractions grouped with an interval of $\Delta Y_\mathrm{e} = 0.05$ are also shown by gray histograms. Note that the recurrence method does not guarantee the physical weights of $\phi_j$'s be unique when the solution degenerates in the $(v/c,\, S,\, Y_\mathrm{e})$-space. It should also be noted that the uncertainties in nuclear ingredients are not considered in this study. Therefore, the distributions of these quantities shown in Figure~\ref{fig:ye} should not be regarded as the unique solution. Keeping these caveats in mind, we find distinct features between mFE-a and mFE-b; the former is represented by $(S,\, Y_\mathrm{e}) \approx (10,\, 0.4)$ and the latter with wide ranges of $(S,\, Y_\mathrm{e})$.

This explains the reason why the abundance distribution of mFE-a extends down to $A = 48$ ($^{48}$Ca). In a condition such as $(S,\, Y_\mathrm{e}) \sim (10,\, 0.4)$, a single NSE cluster contains two maxima at $A = 48$ and $A\approx 64$--84: the former be associated with the magic numbers $(Z, N) = (20, 28)$ and the latter $Z = 28$ or $N = 50$ \citep{Hartmann1985, Meyer1998, Wanajo2018}. Therefore, the production of nuclei with $A = 48$--68 (out of the reference range) is inevitable when the reference abundances contain $A \approx 80$ (the first peak of the residuals). It is noteworthy that the ratio of $^{48}$Ca/Eu (= 751) is comparable (60\%) to that in the solar system \citep[= 1240,][]{Lodders2003}. Given mFE-a be representative, this implies that NS mergers can be potential sources of $^{48}$Ca, whose astrophysical origin is currently unknown \citep{Hartmann1985, Meyer1996, Woosley1997, Wanajo2013, Wanajo2018}.

\section{Heating rates}
\label{sec:heating}

\begin{figure*}
\epsscale{1.17}
\plottwo{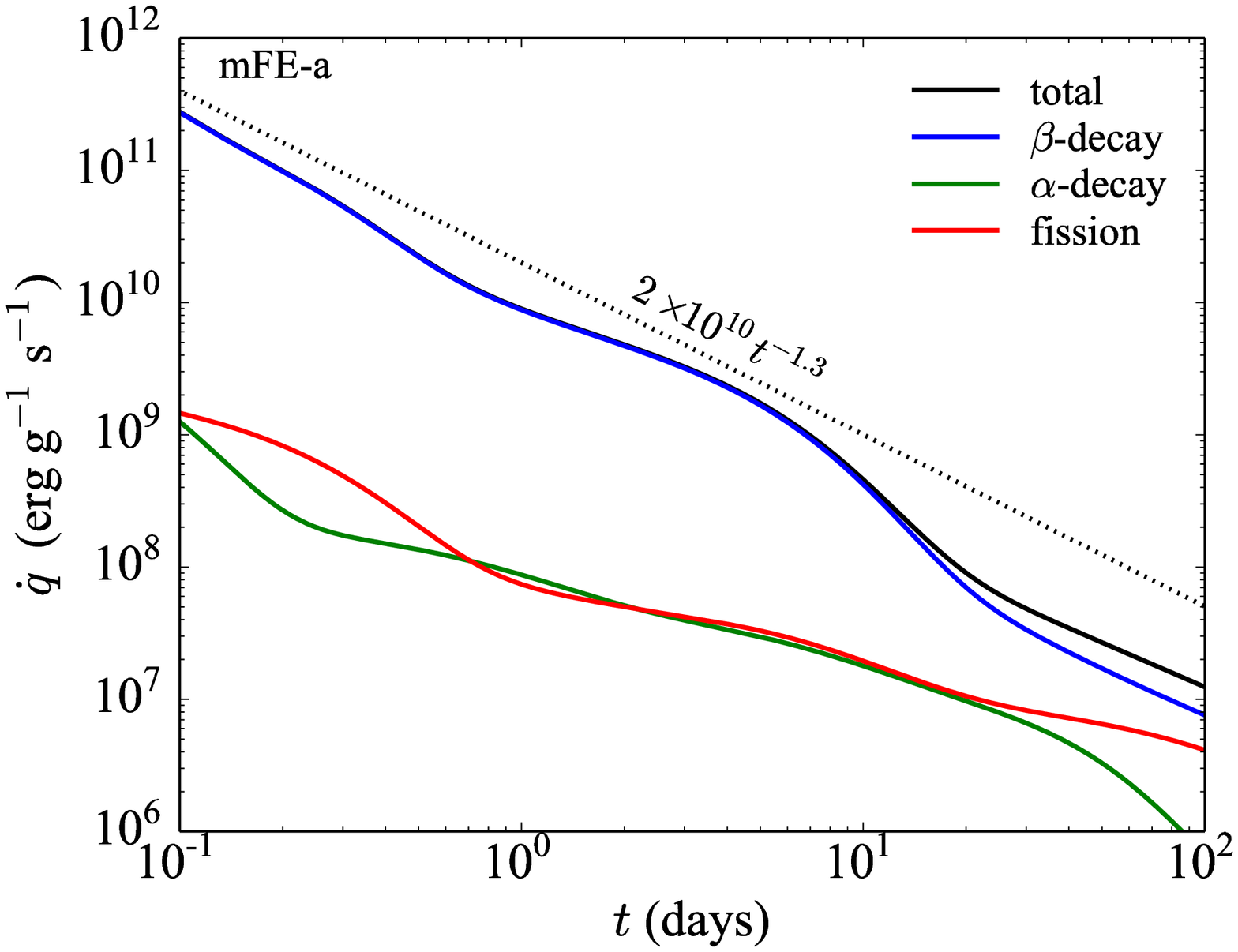}{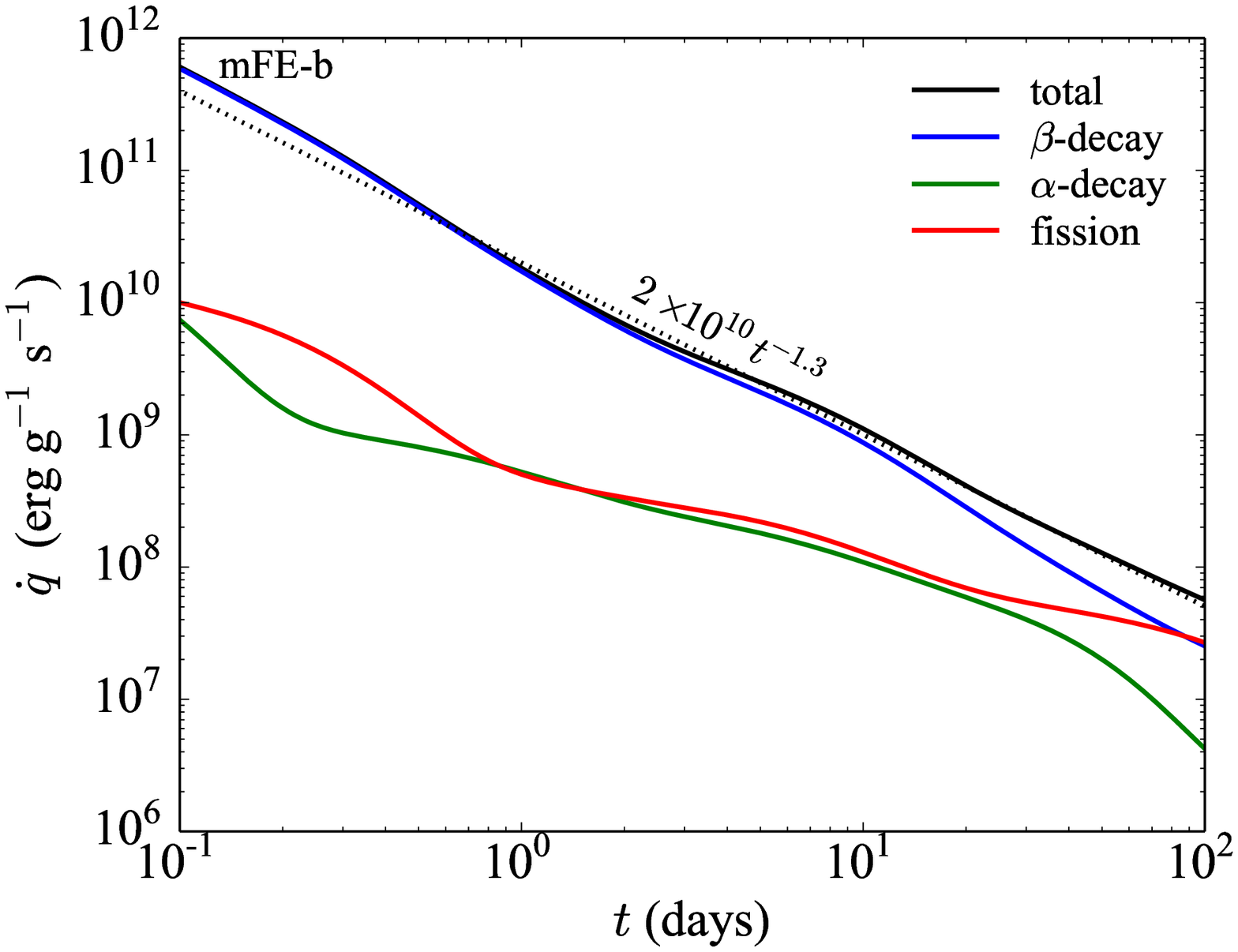}
\plottwo{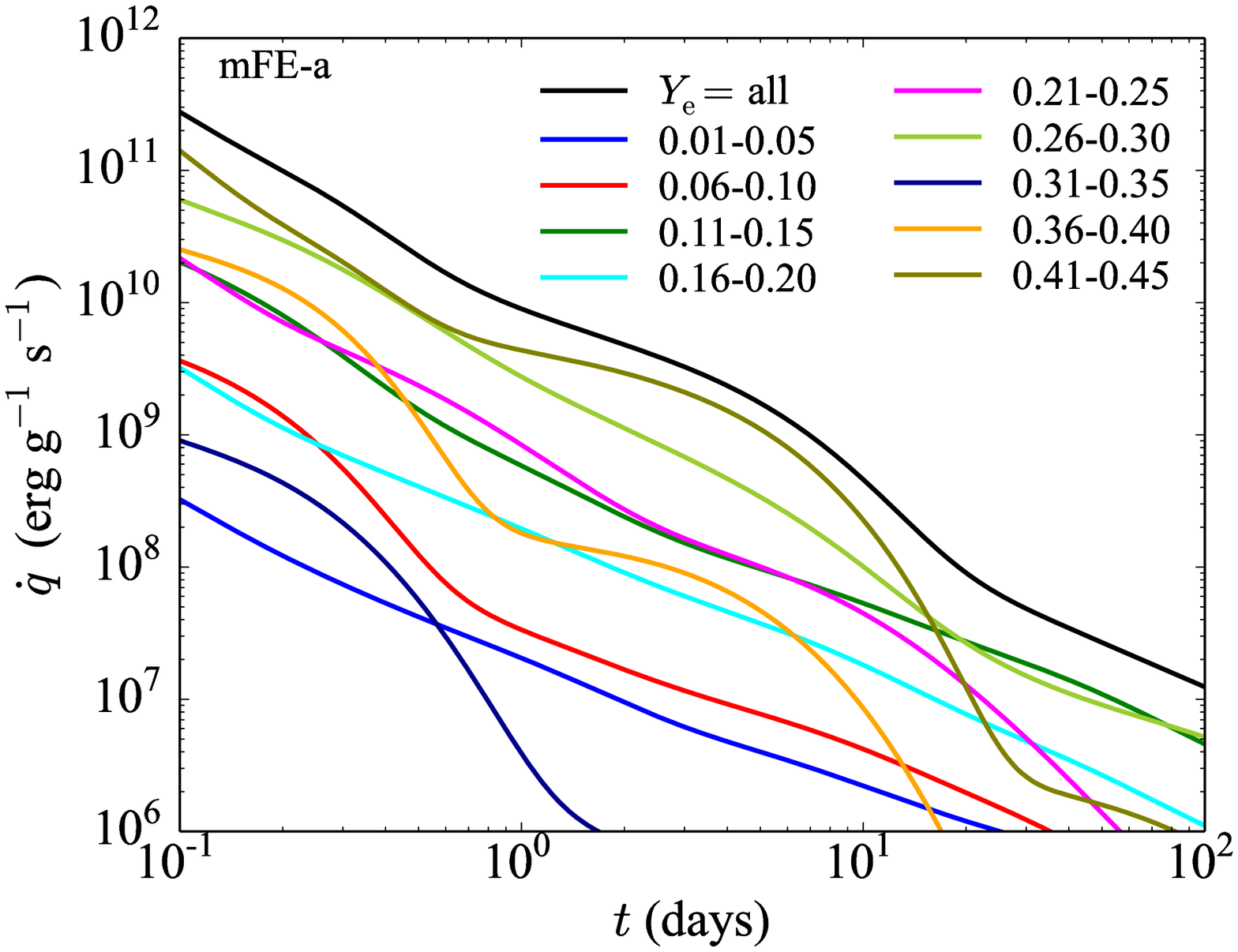}{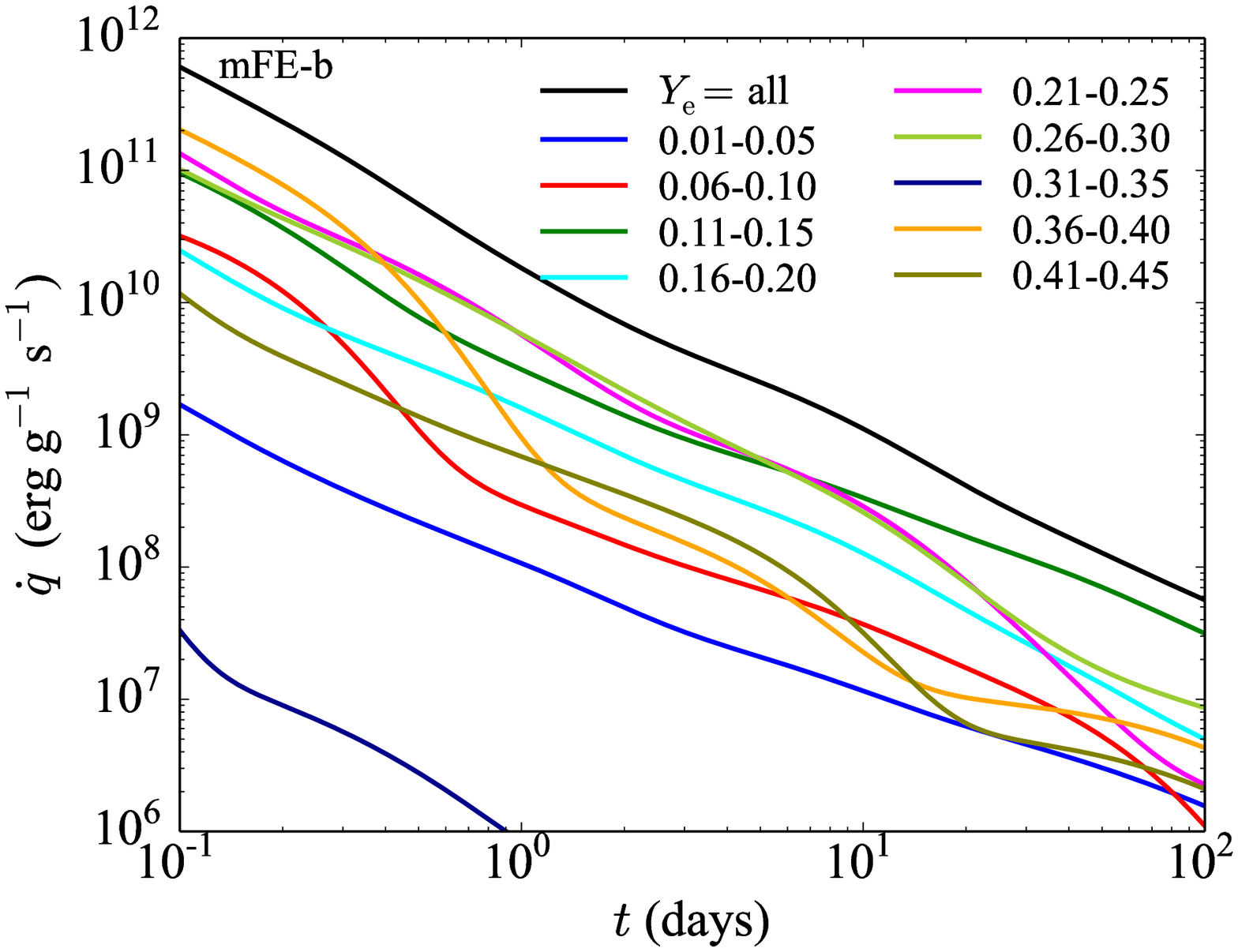}

\caption{\textit{Top}: Heating rates (in erg~g$^{-1}$~s$^{-1}$) for mFE-a (left) and mFE-b (right) as functions of time (in days). The total heating rates are shown in black along with those from $\beta$-decay (blue), $\alpha$-decay (green), and fission (red). Also shown is an empirical law, $2 \times 10^{10}\, t^{-1.3}$~erg~g$^{-1}$~s$^{-1}$ (dotted line). \textit{Bottom}: Heating rates as functions of time (in days) from $Y_\mathrm{e}$ groups (in different colors) along with the total values (black) for mFE-a (left) and mFE-b (right).
}
\label{fig:qdot}
\end{figure*}

\begin{figure*}
\epsscale{1.17}
\plottwo{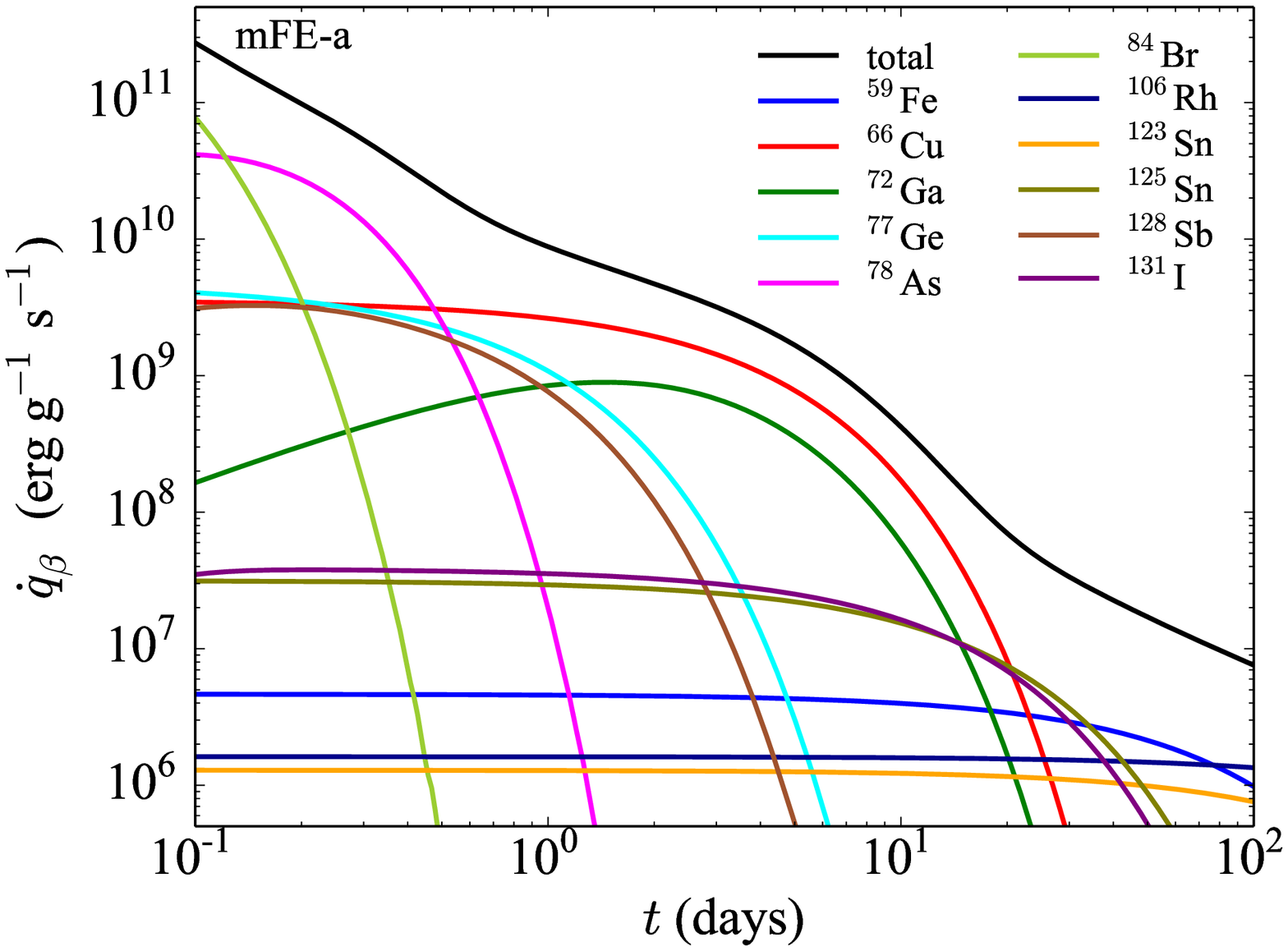}{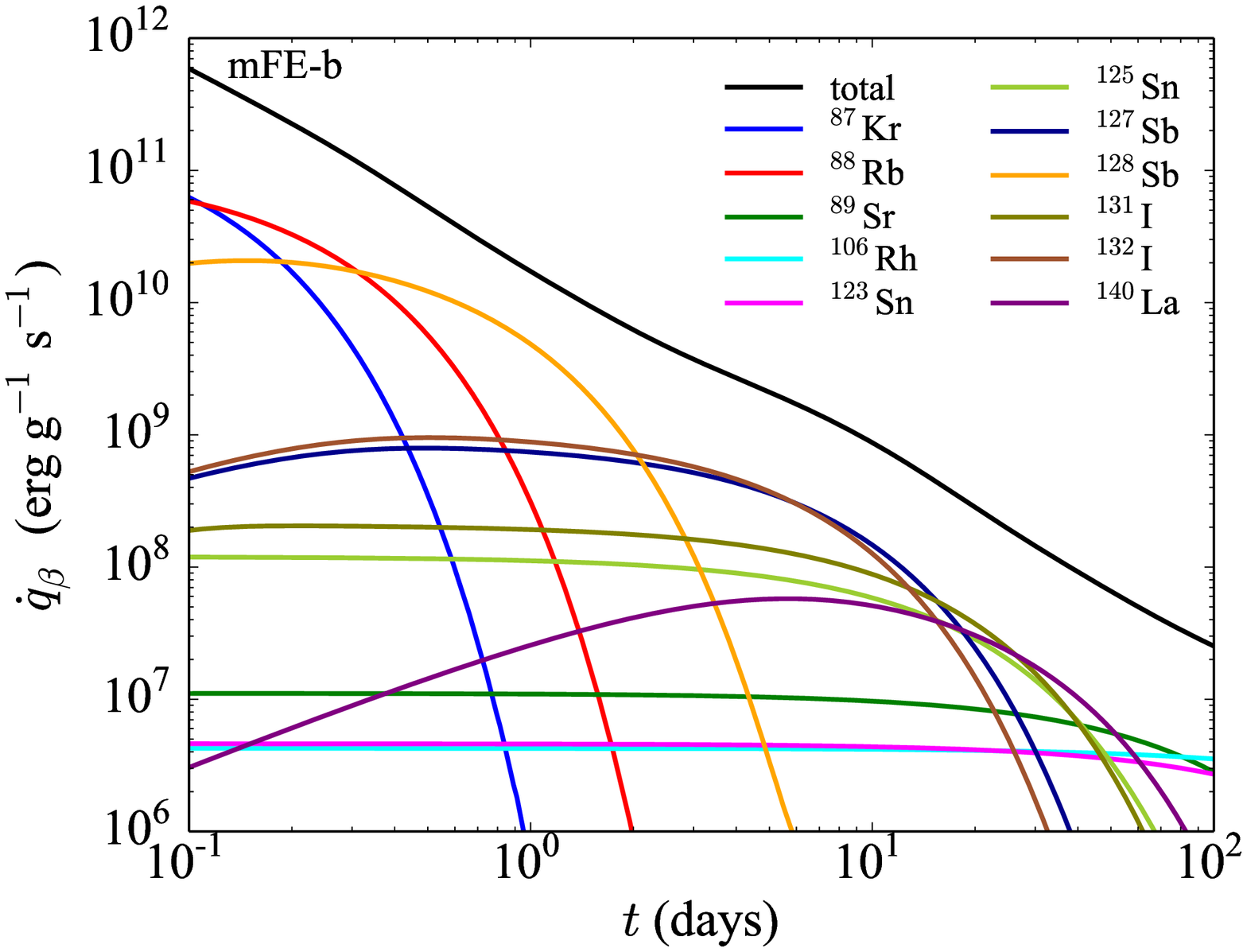}
\plottwo{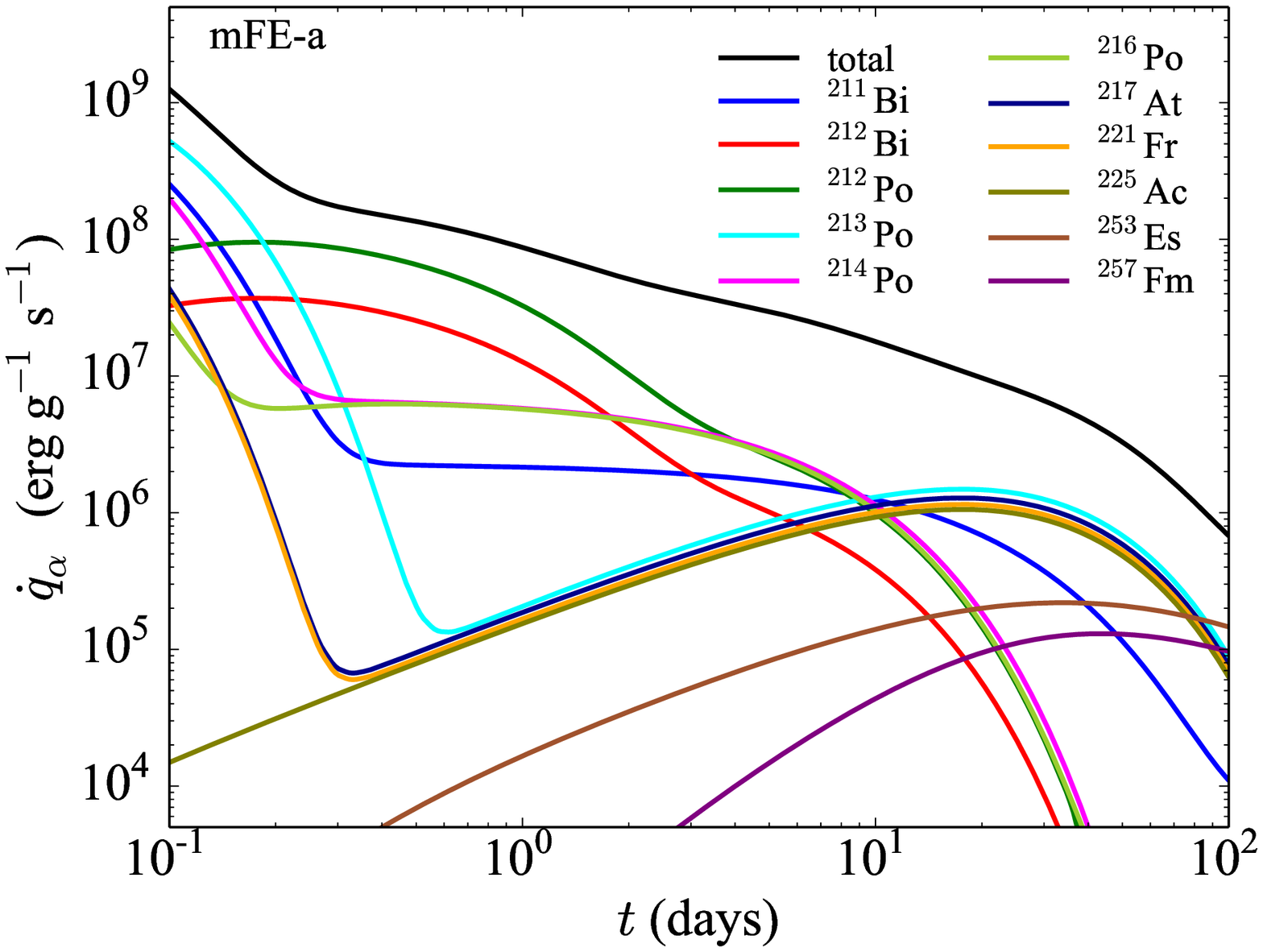}{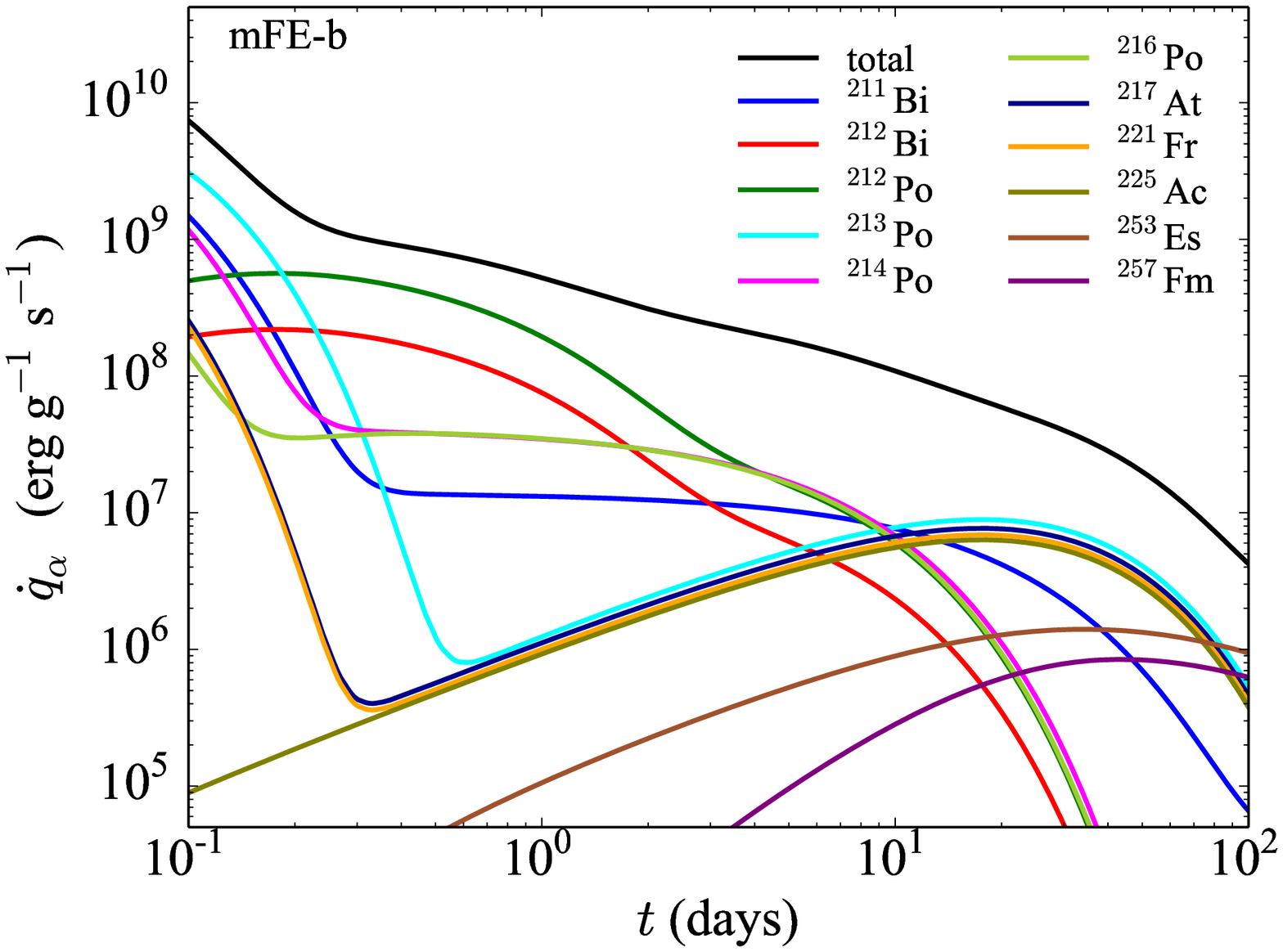}
\plottwo{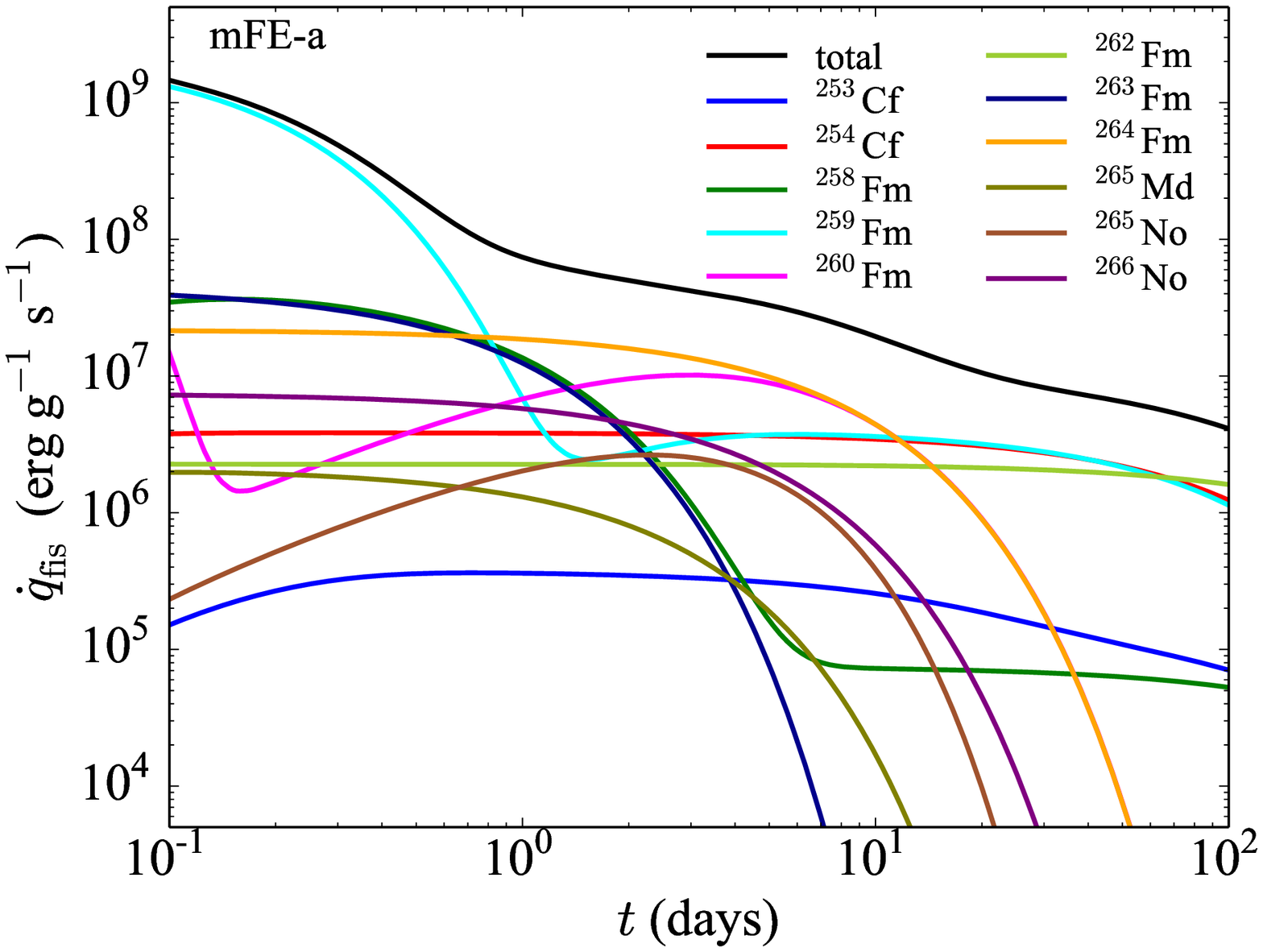}{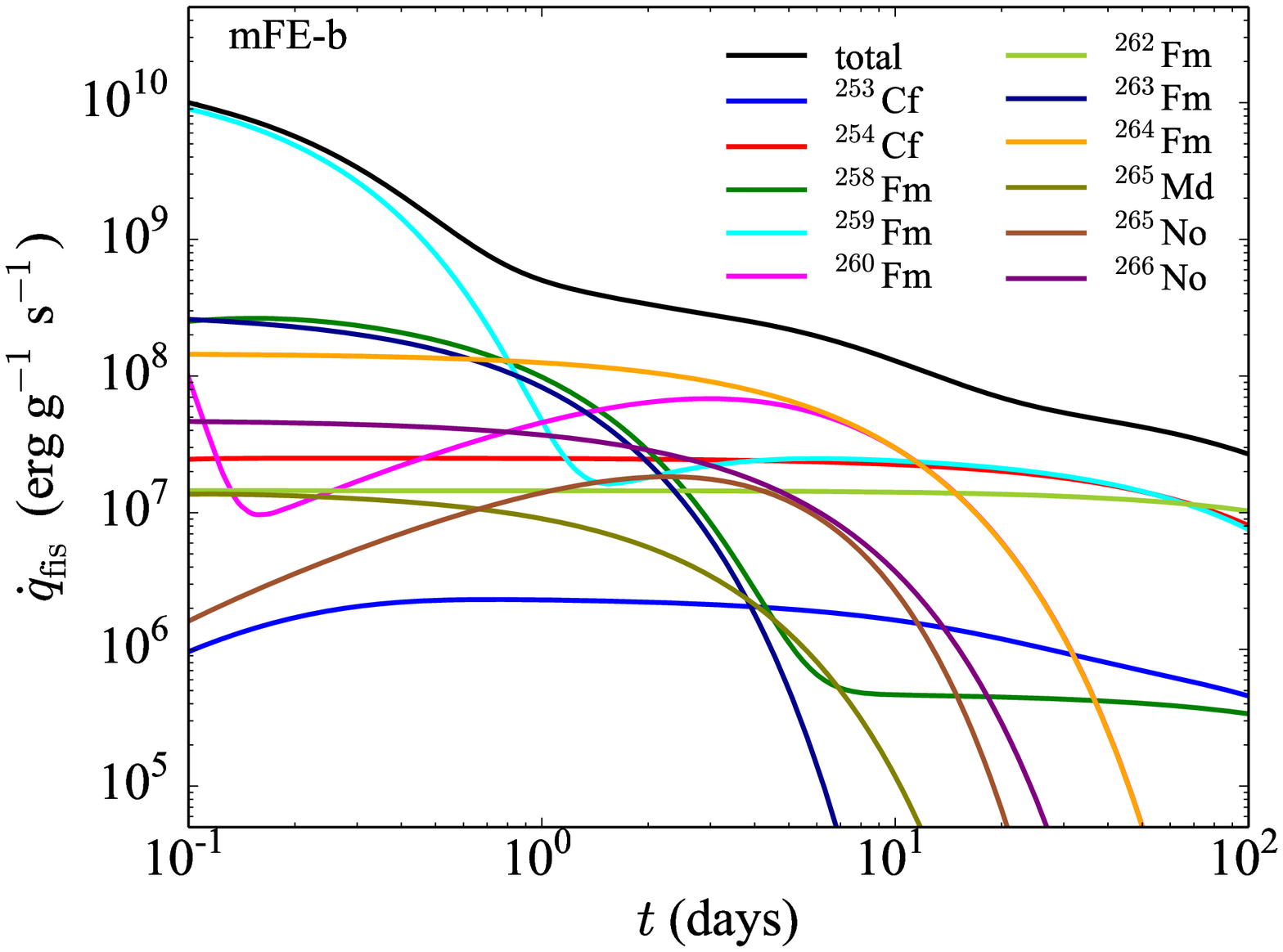}

\caption{Heating rates (black) from $\beta$-decay (top), $\alpha$-decay (middle), and fission (bottom) for mFE-a (left) and mFE-b (right) with the top 11 isotopes (in different colors) that have more than 10\% contributions at the maxima.
}
\label{fig:qdot_isotope}
\end{figure*}

An ensemble of FEs with their weights gives the temporal evolution of the nuclear heating rates for each mFE,
\begin{equation}
    \label{eq:qdot}
    \dot{q}(t) = \sum_{j=1}^{N_\mathrm{FE}} \phi_j \, \dot{q}_{\mathrm{FE}, j} (t),
\end{equation}
where $\dot{q}_{\mathrm{FE}, j} (t)$ is the heating rate of the $j$th FE as a function of time. The resulting heating rates are shown in Figure~\ref{fig:qdot} for mFE-a (left) and mFE-b (right) with those from all channels (black), $\beta$-decay (blue), $\alpha$-decay (green), and fission (red) as functions of time\footnote{Numerical data for the upper panels of Figure~\ref{fig:qdot} are available. See also \\ https://sites.google.com/view/shinyawanajo/}. For both mFEs, the contribution from $\beta$-decay dominate over the others; fission and $\alpha$-decay become important at late times \citep{Hotokezaka2016, Barnes2016}. While the total heating rate in mFE-b well scales as $\approx 2 \times 10^{10}\, t^{-1.3}$~erg~g$^{-1}$~s$^{-1}$ \citep[e.g.,][]{Metzger2010, Wanajo2014}, the same does not hold in mFE-a. The behaviors of $\alpha$-decay and fission contributions are similar between mFE-a and mFE-b but with appreciably greater values for the latter (because of the greater amount of $r$-process elements). In mFE-a (Figure~\ref{fig:qdot}, bottom left) the FEs of $Y_\mathrm{e} = 0.41$--0.45 dominate over those of the other $Y_\mathrm{e}$ ranges, while in mFE-b (right) the $Y_\mathrm{e}$ groups of 0.06--0.35 have similar contributions. This is due to the dominance of FEs with $Y_\mathrm{e} = 0.41$ in mFE-a (Figure~\ref{fig:ye}).

The reason of deviation from a power law in mFE-a is attributed to the two $\beta$-decay chains that principally contribute to heating (halflife),
\begin{eqnarray}
    \label{eq:a66}
    ^{66}\mathrm{Ni}\, (2.28\,\mathrm{d}) & \rightarrow ^{66}\mathrm{Cu}\, (5.12\,\mathrm{m}) & \rightarrow ^{66}\mathrm{Zn}, \\
    \label{eq:a72}
    ^{72}\mathrm{Zn}\, (1.94\,\mathrm{d}) & \rightarrow ^{72}\mathrm{Ga}\, (14.1\,\mathrm{h}) & \rightarrow ^{72}\mathrm{Ge},
\end{eqnarray}
as can be seen in Figure~\ref{fig:qdot_isotope} (top left). The parent isotopes $^{66}$Ni and $^{72}$Zn are made in NSE. The decay chain from $^{66}$Ni exhibits a factor of a few greater contribution than that from $^{72}$Zn. $A = 66$ is out of the range from the residuals but co-produced with $^{72}$Zn in the same condition $(S,\, Y_\mathrm{e}) \approx (10,\, 0.4)$ as described in \S~\ref{sec:reference}. The dominance of only two isobars $A = 66$ and 72 with similar halflives of the parent nuclides $^{66}$Ni and $^{72}$Zn leads to an exponential-like (rather than power-law-like) behavior of the heating rate during $t = 1$--15 days. Note that, in mFE-a, the isotopes of $A \sim 130$ play only subdominant roles on the $\beta$-decay heating. In mFE-b (Figure~\ref{fig:qdot_isotope}, top right) those of $A \sim 130$ have dominant contributions, exhibiting a power-law-like decay as found in previous studies \citep[e.g.,][]{Metzger2010, Wanajo2014}.

Dominant contributing isotopes for $\alpha$-decay and fission are also shown in Figure~\ref{fig:qdot_isotope} (middle and bottom). As anticipated, little difference can be seen between mFE-a (left) and mFE-b (right) except for consistently greater values in the latter. For fission, $^{254}$Cf and a few Fm isotopes mainly contribute to the heating rate \citep{Wanajo2014, Zhu2018}. Note that the measured halflives of $^{258, 259, 260}$Fm for spontaneous fission are only 370~$\mu$s, 1.5~s, and 4~ms, respectively, but sizably contribute to heating because of predicted longer $\beta$-decay lifetimes of the parent isotopes adopted in this study. As the contributions of these isotopes are all subject to very uncertain $\beta$-decay and spontaneous fission lifetimes of their parent nuclides, the outcomes should be taken at a qualitative level.

\section{Implication to kilonova light curves}
\label{sec:kilonova}

\begin{deluxetable}{cccccc}
\tabletypesize{\scriptsize}
\tablecaption{Ejecta masses ($M_\odot$) for GW170817}
\tablewidth{0pt}
\tablehead{
\colhead{Model} &
\colhead{total} &
\colhead{$A \ge 90$} &
\colhead{$A \ge 139$} &
\colhead{Eu}
}
\startdata
mFE-a & 0.06 & 0.088 & 0.0021 & $2.6\times 10^{-5}$ &  \\ 
mFE-b & 0.04 & 0.029 & 0.083  & $1.1\times 10^{-4}$ &  \\ 
\enddata
\label{tab:ejecta}
\end{deluxetable}

\begin{deluxetable}{ccccc}
\tabletypesize{\scriptsize}
\tablecaption{Ratios of Th/Eu in $r$-process-enhanced stars}
\tablewidth{0pt}
\tablehead{
\colhead{Star} &
\colhead{now\tablenotemark{a}} &
\colhead{10 Gyr ago\tablenotemark{b}} &
\colhead{13.5 Gyr ago\tablenotemark{c}}
}
\startdata
J09544277+5246414 & 0.76 & 1.2 & 1.5 &  \\ 
DES~J033523-540407 & 0.14 & 0.24 & 0.28 &   \\ 
\enddata
\tablenotetext{a}{Measured value.}
\tablenotetext{b}{Corrected value (10~Gyr ago).}
\tablenotetext{c}{Corrected value (13.5~Gyr ago).}
\label{tab:thorium}
\end{deluxetable}

\begin{figure*}
\epsscale{1.17}
\plottwo{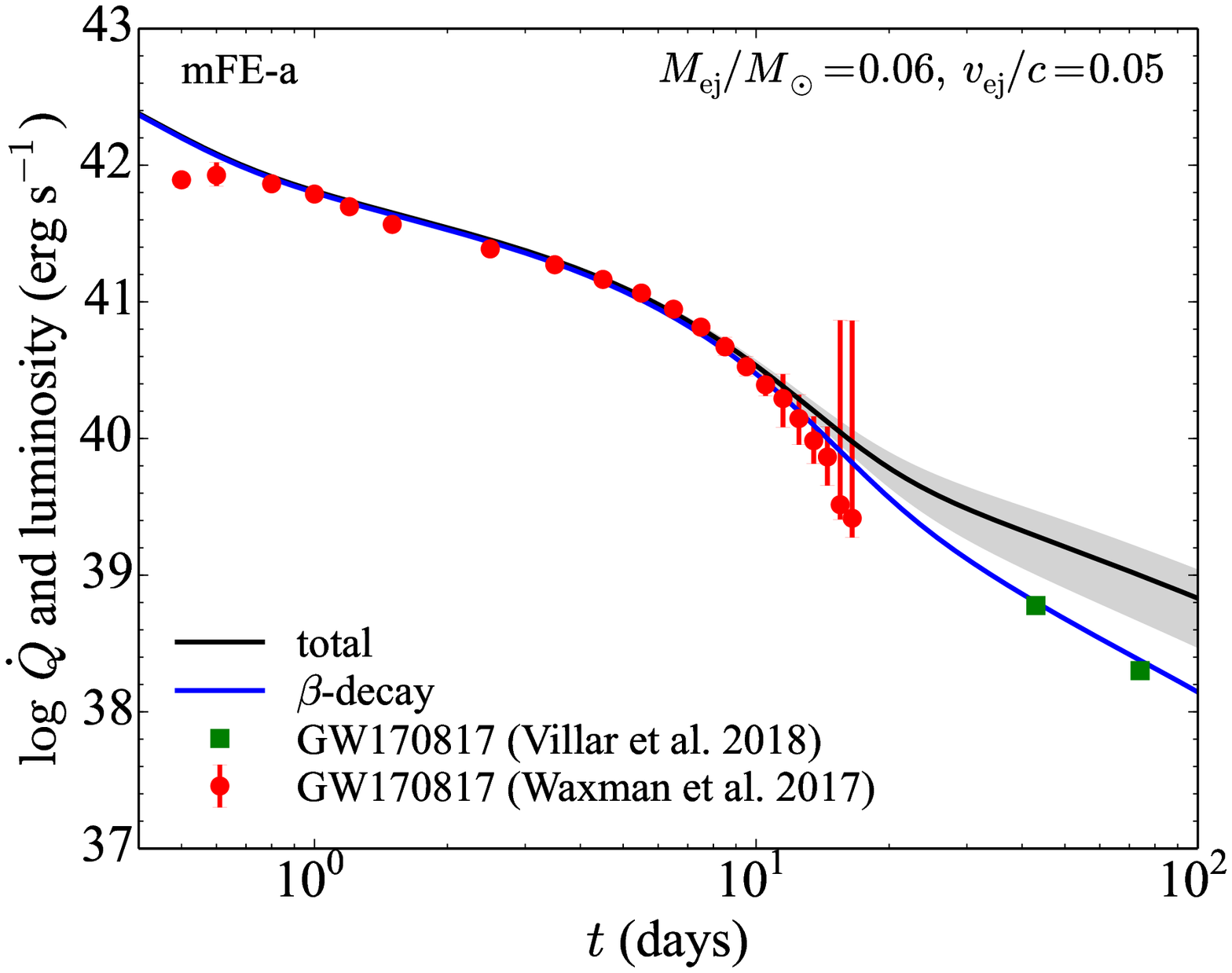}{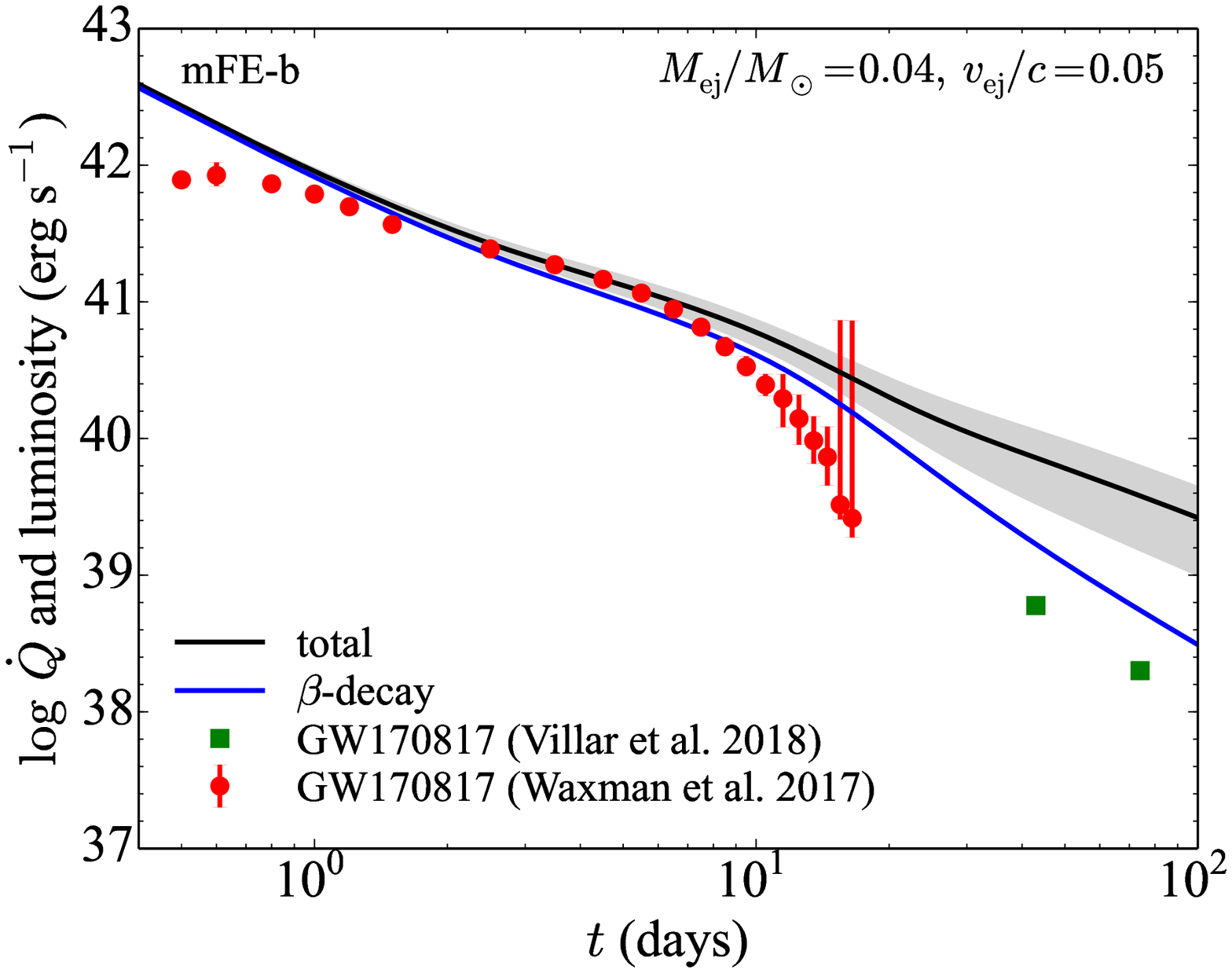}
\plottwo{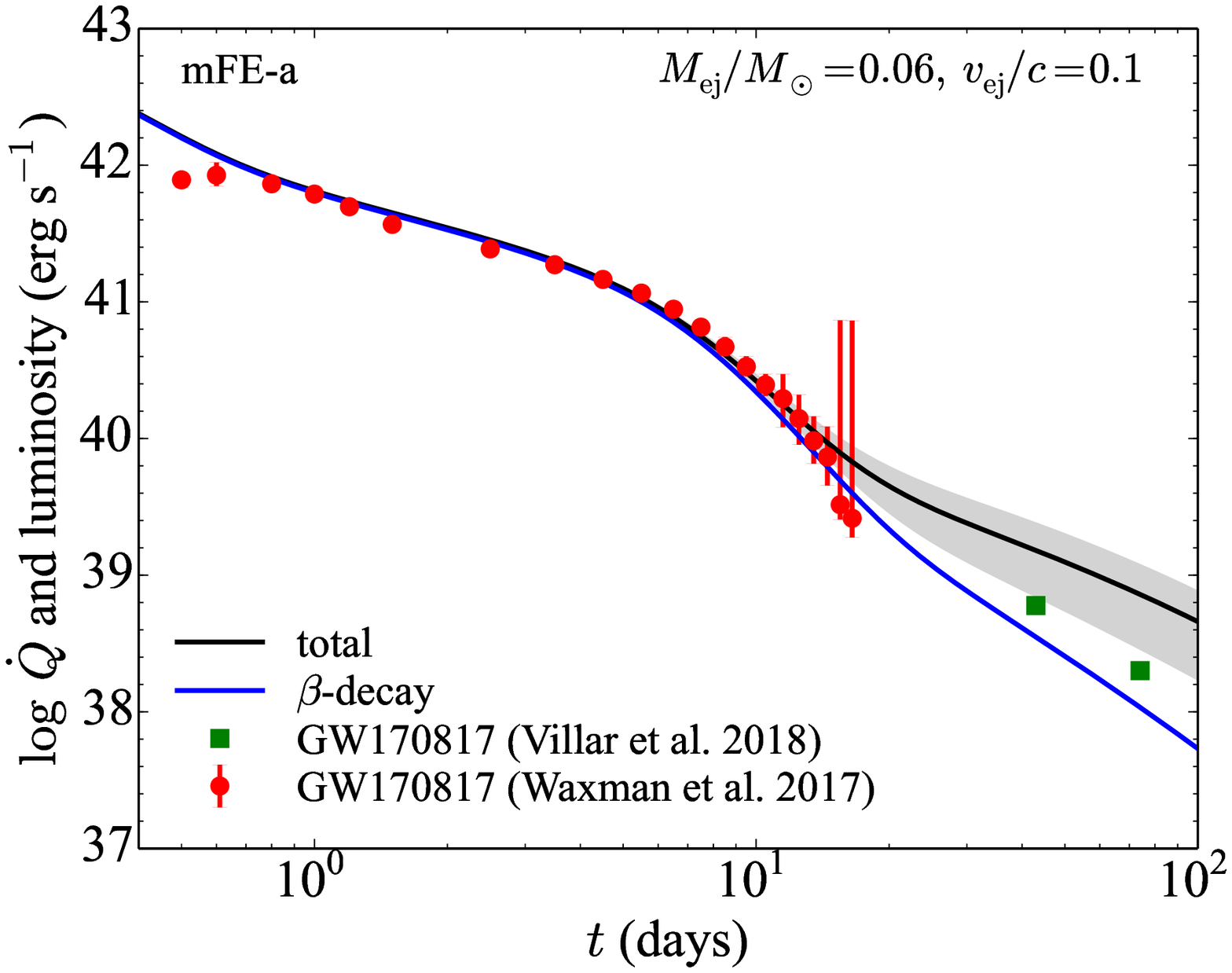}{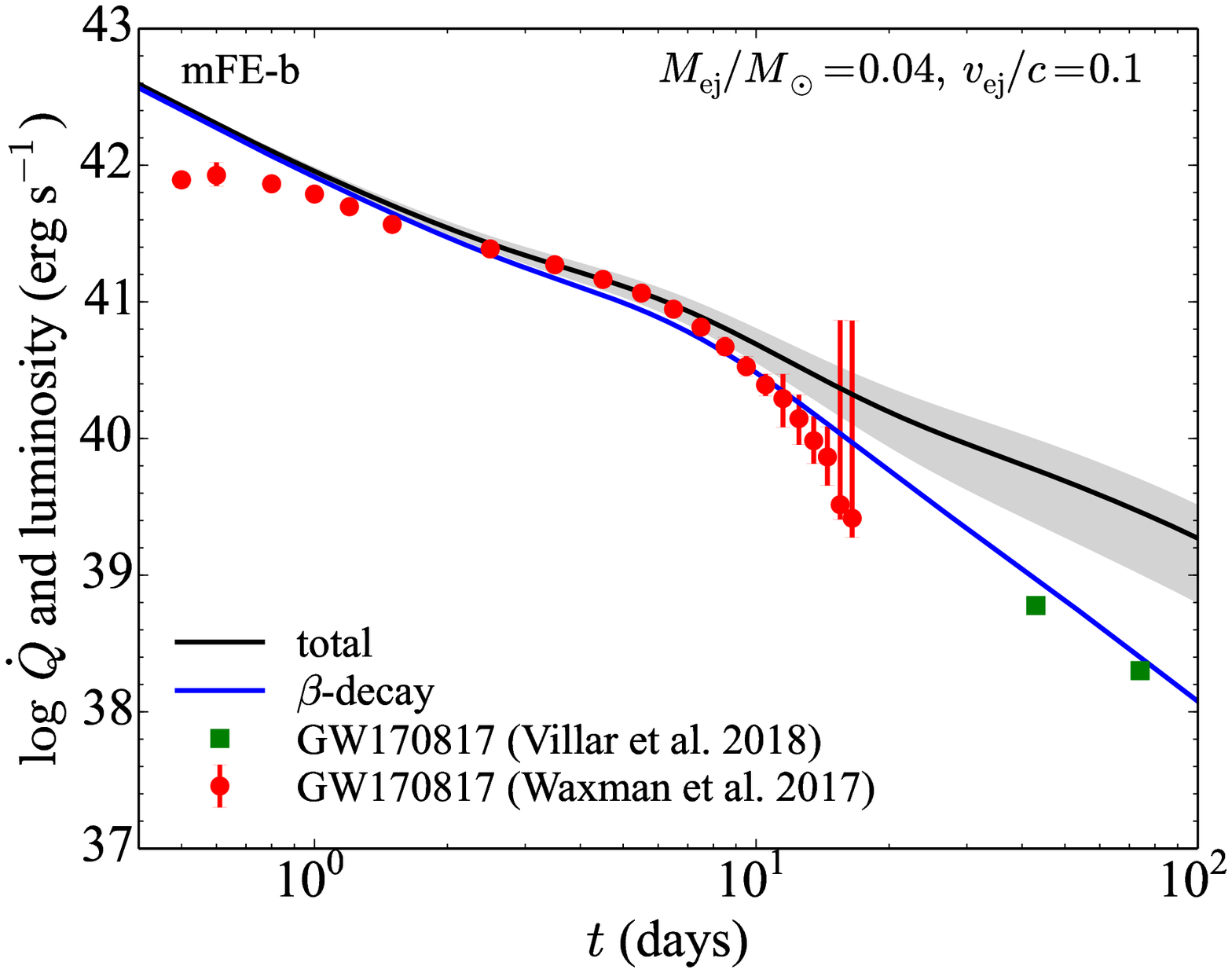}
\plottwo{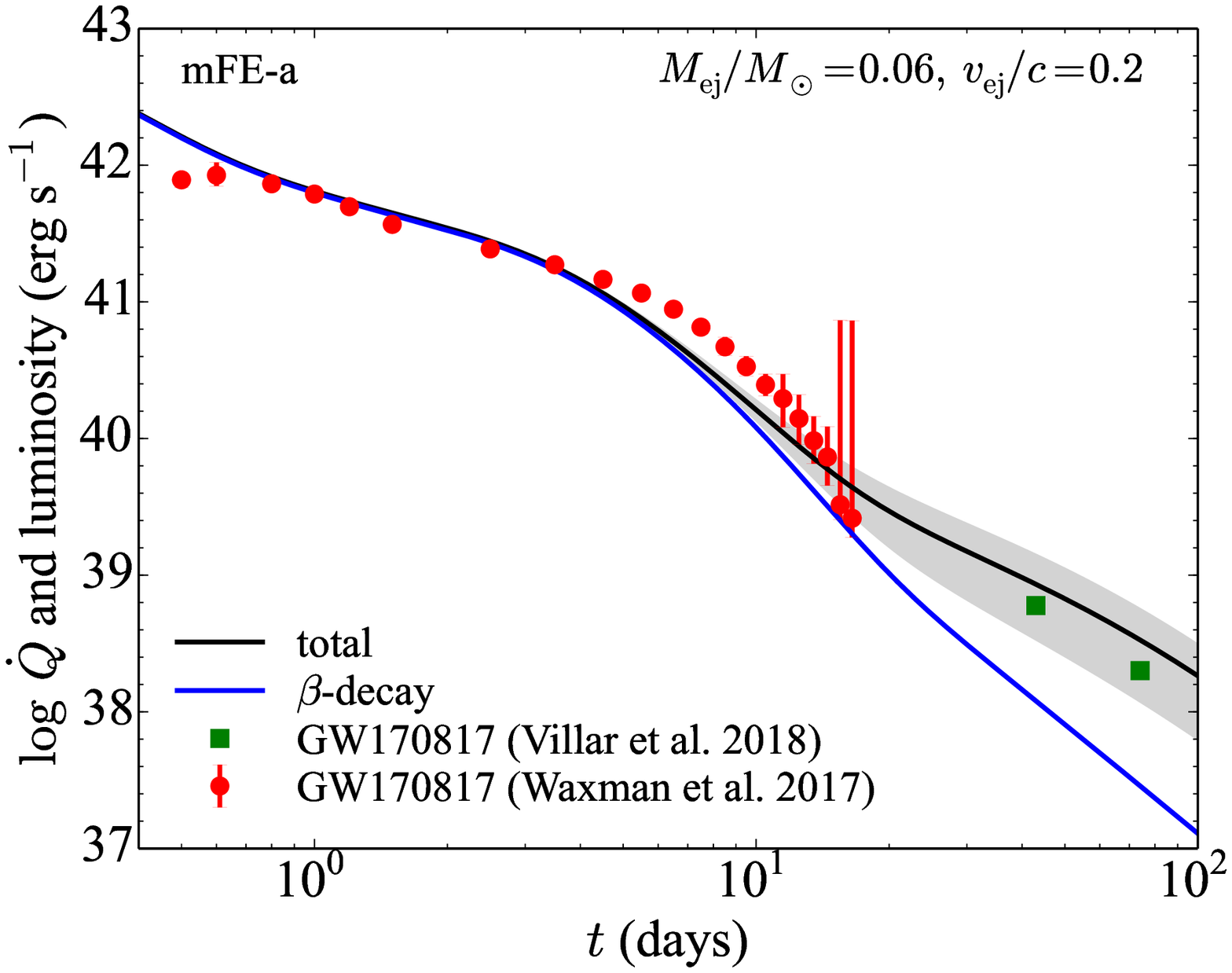}{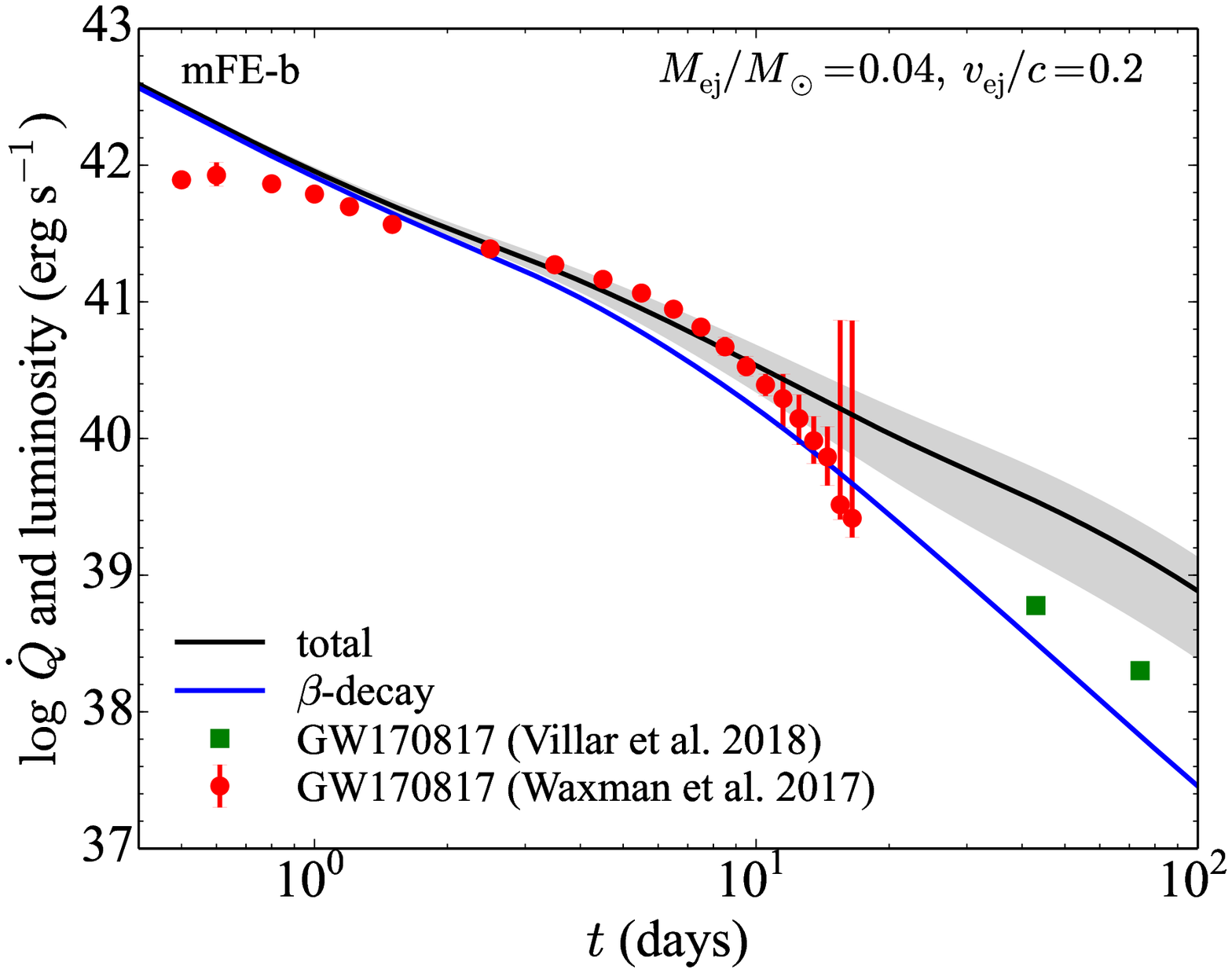}

\caption{Comparison of the heating rate (in units of erg~s$^{-1}$) with the ejecta mass $M_\mathrm{ej}$ and velocity $v_\mathrm{ej}$ (denoted in the legend) with the bolometric luminosity of the kilonova (AT~2017gfo or SSS17a) associated with the NS merger GW170817 adopted from \citet[][red circles with error bars]{Waxman2017} and \citet[][green squares without error bars]{Villar2018}. The total and $\beta$-decay heating rates are displayed by black and blue lines, respectively. The latter can be regarded as the lower bound when considering uncertainties in the fission contribution. The grey area indicates the range inferred from a variation of Th/Eu in $r$-process-enhanced stars.
}
\label{fig:luminosity}
\end{figure*}

Provided that the light curve after a few days since merger is exclusively due to radioactive energies, we compare the heating rates obtained in \S~\ref{sec:heating} with the bolometric luminosity of the kilonova (AT~2017gfo or SSS17a) associated with the NS merger GW170817. The heating rate in the ejecta (in erg~s$^{-1}$) is calculated as
\begin{equation}
    \dot{Q}(t) = M_\mathrm{ej}\, \epsilon(t)\, \dot{q}(t), 
\end{equation}
where the thermalization efficiency $\epsilon(t)$ for given ejecta mass $M_\mathrm{ej}$ and velocity $v_\mathrm{ej}$ is obtained from the analytic formula in \citet[][Eqs.~16, 21, and 25]{Barnes2016}. Note that $v_\mathrm{ej}$ is taken as an independent free parameter from $v$ in FEs, because mFE models do not necessarily give unique solutions of the velocity distributions as described in \S~\ref{sec:reference}. The adopted energy partitions of $\beta$-decay are 0.4 to neutrinos (that do not contribute to heating), 0.4 to electrons, and 0.2 to $\gamma$-rays according to \citet[][Figure~1]{Hotokezaka2016}.

$\dot{Q}$ for each mFE is compared with the observed bolometric luminosity of the kilonova associated with GW170817 \citep{Waxman2017, Villar2018}. For mFE-a and mFE-b, the ejecta masses are taken to be $M_\mathrm{ej}/M_\odot = 0.06$ and 0.04, respectively (Table~2, second column), such that $\dot{Q}$ approximately matches the light curve at a few days since merger. A greater $M_\mathrm{ej}$ for the former is needed to compensate the smaller heating rate per mass (Figure~\ref{fig:qdot}, top). Three different ejecta velocities $v_\mathrm{ej}/c = 0.05$ (top), 0.10 (middle), and 0.20 (bottom) are considered, which affect the thermalization efficiency $\epsilon(t)$. The calculated total heating rate $\dot{Q}(t)$ and its $\beta$-decay component are displayed by black and blue lines, respectively. The latter can be regarded as the lower bound of the heating rate when considering uncertainties in the fission contribution described in \S~\ref{sec:reference}.

On top of the uncertainties originating from nuclear ingredients, an intrinsic  variation of actinide productivity is known from the measurements of Th/Eu ratios in $r$-process-enhanced stars, which can be taken as the range of Th/Eu production ratio. Among those, a recently found Galactic halo star, J09544277+5246414 \citep{Holmbeck2018}, exhibits the highest ratio of  Th/Eu = 0.76  (Table~\ref{tab:thorium}, second column). Given the age of the oldest Galactic halo star be 13.5~Gyr, the upper limit of the production ratio becomes Th/Eu = 1.5 (Table~\ref{tab:thorium}, 4th column) with  the halflife of $^{232}$Th (14.05~Gyr). The lowest measured Th/Eu (= 0.14) has been reported for one of $r$-process-enhanced stars, DES~J033523-540407 \citep{Ji2018}, in a ultra-faint dwarf galaxy Reticulum~II. Assuming the age of the youngest star in Reticulum~II be 10~Gyr, the lower limit becomes 0.28. Our result indicates the Th/Eu production ratio (Table~\ref{tab:properties}, 7th column) in between the inferred range above (Th/Eu = 0.28--1.5). Given the amount of trans-lead nuclei be proportional to Th/Eu, the range of the heating rate with rescaled contributions from fission and $\alpha$-decay is indicated by the grey area in Figure~\ref{fig:luminosity}.

Obviously, the light curve of GW170817 is best explained by mFE-a, in particular with $v_\mathrm{ej}/c = 0.1$ (middle left), for both a steepening at $t \gtrsim 7$~days \citep{Waxman2017} and the late-time estimates \citep[43 and 74~days,][]{Villar2018}. The former is indicative of the two $\beta$-decay chains from $^{66}$Ni and $^{72}$Zn in Eqs.~(\ref{eq:a66})-(\ref{eq:a72}) with similar halflives ($\approx 2$~days). Note that a slower ejecta velocity ($v_\mathrm{ej}/c = 0.05$, top left) keeps a high thermalization efficiency \citep{Barnes2016} and thus $\dot{Q}(t)$ becomes slightly less steeper, which is however still consistent within the error bars. For $v_\mathrm{ej}/c = 0.2$, $\dot{Q}$ becomes too steep to be consistent with the light curve. At late times for $v_\mathrm{ej}/c = 0.1$, the data points are marginally consistent with the lower limit of the heating rate (in grey) and slightly greater than that from $\beta$-decay (i.e., the lower bound). Note that the ejecta might be in a nebular phase at these late times, in which the thermalization factor $\epsilon(t)$  could be overestimated.

It appears a reasonable interpretation, therefore, that the products from the NS merger GW170817 consist of mainly light trans-iron elements ($A < 90$) with a small fraction of $r$-process elements ($A \ge 90$). This also explains an inferred small content of lanthanides and heavier in the ejecta ($X_\mathrm{l} \lesssim 0.01$). Assuming that mFE-a with $M_\mathrm{ej}/M_\odot = 0.06$ represents the merger GW170817, the ejecta mass of the $r$-process elements is $\approx 0.01\, M_\odot$ (Table~2, 3rd column), which is fully consistent with that from the dynamical ejecta of a NS merger \citep[e.g.,][]{Shibata2017}. The rest of $\approx 0.05\, M_\odot$ that consists of light trans-iron elements can also be explained by the post-merger disk outflows in \citet{Fujibayashi2018}. In their model of $\alpha_\mathrm{vis} = 0.04$, the ejecta mass and mean velocity are, respectively, $\approx 0.05\, M_\odot$ and $\approx 0.1\, c$ (in their Figure~8), which coincide with those of our best case in mFE-a (Figure~\ref{fig:luminosity}, middle left). In their relevant model the distributions of $S$ and $Y_\mathrm{e}$ exhibit relatively narrow peaks at $\approx 10\, k_\mathrm{B}/\mathrm{nuc}$ and $\approx 0.35$--0.4, respectively (their Figure~14), which are  (marginally for the latter) in agreement with those in mFE-a (Figure~\ref{fig:ye}). An updated analysis of \citet{Metzger2014} in \citet{Lippuner2017} also shows similar trends. The mass of Eu estimated for GW170817 is $2.6\times 10^{-5}\, M_\odot$ (Table~2, 5th column) that also is in agreement with the values adopted in recent works of Galactic chemical evolution \citep[Figure~1 in][]{Cote2017}.

For mFE-b (Figure~\ref{fig:luminosity}, right), none of cases accounts for the steepening at $t \gtrsim 7$~days. Effects of velocity in $\epsilon(t)$ are too small to steepen the light curve. The late-time heating rates also are largely over-predicted except for the fast velocity case ($v_\mathrm{ej}/c = 0.2$). We cannot exclude a possibility that the ejecta quickly became transparent at $t \gtrsim 7$~days and a reduction of thermalization resulted in the steepening and smaller luminosities at late times. Otherwise, the bolometric luminosities at late times might be largely overestimated. However, a tension between observations and theoretical results stands; the $r$-process products as massive as $\approx 0.5\, M_\odot$ cannot be explained neither by dynamical ejecta or disk outflows. In addition, the inferred mass fraction of lanthanides and heavier ($\lesssim 0.01$) disagree with that in mFE-b ($\approx 0.2$; Table~\ref{tab:properties}, 4th column).

\section{Summary and conclusions}
\label{sec:summary}

Multi-component free-expansion model (mFE) was constructed to investigate various aspects of physical conditions relevant for the $r$-process. This paper was supposed to be the first of a series of papers with an emphasis on radioactive heating rates that power a kilonova, an electromagnetic counterpart of the gravitational signals from a NS merger. Each free expansion model (FE) consists of a three-parameter suite: expansion velocity $v/c\, (= 0.05$--0.30), entropy $S\, (= 10$--35 in units of $k_\mathrm{B}/\mathrm{nuc}$), and electron fraction $Y_\mathrm{e}\, (= 0.01$--0.50). A mFE was defined as an ensemble of FEs that fitted a given reference abundance distribution. The $r$-process residuals (by subtracting the $s$-process component) to the solar system abundances \citep{Goriely1999} was employed as the reference, in light of the robust (solar-$r$-process like) abundance patterns in $r$-process-enhanced stars \citep[at least in the range $50 < Z < 80$, e.g.,][]{Sneden2008}.

Two models were considered: a) mFE-a with a reference of the full range of the residuals ($A \ge 69$) with both light trans-iron and $r$-process elements and b) mFE-b with $r$-process elements only ($A \ge 90$). For mFE-b, the fitting of FEs to the reference resulted in wide ranges of $S$ and $Y_\mathrm{e}$ as found in numerical simulations of dynamical ejecta (\citealt{Wanajo2014, Sekiguchi2015, Sekiguchi2016}; see also a similar trend in the magneto-hydrodynamic simulations of accretion disks by \citealt{Siegel2017, Fernandez2018}). In contrast, mFE-a composed of narrow ranges of $S\approx 10$ and $Y_\mathrm{e} \approx 0.4$ that were in good agreement with a recent simulation of disk outflows (\citealt{Fujibayashi2018}; see also \citealt{Lippuner2017}). As such physical conditions led to NSE, the nucleosynhthetic yields extended down to $^{48}$Ca, a nuclide whose astrophysical origin was unknown \citep{Hartmann1985, Meyer1996, Woosley1997, Wanajo2013b}.

While the obtained heating rate for mFE-b exhibited a power-law-type temporal evolution owing to $\beta$-decay (of $A\sim 130$) as found in previous works \citep[$\approx 2 \times 10^{10}\, t^{-1.3}$~erg~g$^{-1}$~s$^{-1}$, e.g.,][]{Metzger2010, Wanajo2014}, that for mFE-a indicated rather an exponential-type evolution during $t\approx 1$--15~days. Two $\beta$-decay chains relevant for the heating were identified: $^{66}$Ni$\rightarrow^{66}$Cu$\rightarrow^{66}$Zn and $^{72}$Zn$\rightarrow^{72}$Ga$\rightarrow^{72}$Ge with both having similar halflives ($\approx 2$~days) of the parent isotopes made in NSE. Contributions from fission and $\alpha$-decay became non-negligible at late times ($> 10$~days) as pointed out by \citet{Hotokezaka2016, Barnes2016, Zhu2018}.

Obtained heating rates with given ejecta mass $M_\mathrm{ej}$ and ejecta velocity $v_\mathrm{ej}$ \citep[both affect the thermalization efficiency,][]{Barnes2016} were compared with the bolometric light curve of the kilonova (AT~2017gfo or SSS17a) associated with GW170817. It was found that the light trans-iron dominant model mFE-a with $M_\mathrm{ej}/M_\odot = 0.06$ and $v_\mathrm{ej}/c = 0.1$ reproduced the bolometric light curve remarkably well. A steepening of the light curve at $\gtrsim 7$~days \citep{Waxman2017} was indicative of the dominance of light trans-iron nuclei including $^{66}$Ni and $^{72}$Zn, rather than $r$-process products. Late-time estimates at several 10 days \citep{Villar2018} could also be explained with the $\beta$-decay heating and probably those from fission and $\alpha$-decay. The pure $r$-process model (mFE-b with $M_\mathrm{ej}/M_\odot = 0.04$) did not account for the light curve at late times because of its robust power-law like decay of the heating rate. Note that our models account for the gross nature of the kilonova light curve but its early blue component that may originate from a high-latitude dynamical ejecta \citep[e.g.,][]{Shibata2017} or the wind ejecta from a strongly magnetized hypermassive NS \citep{Metzger2018}.

In conclusion, the ejecta from the NS merger GW170817 was dominated ($\approx 0.05\, M_\odot$) by light trans-iron elements ($A < 90$) with a fraction ($\approx 0.01\, M_\odot$) of $r$-process elements ($A \ge 90$). Along with the adopted velocity ($v_\mathrm{ej}/c = 0.1$), our conclusion is consistent with an interpretation that the $r$-process elements come from the dynamical ejecta of a NS merger and light trans-iron elements from the subsequent disk outflows \citep[e.g.,][]{Shibata2017}. Magneto-hydrodynamic mass ejection from an accretion disk may also be a viable mechanism for this event \citep{Siegel2017, Fernandez2018}. Although the inferred $M_\mathrm{ej}$ and $v_\mathrm{ej}$ are similar to the literature values, it is emphasized that the principal radioactive energy sources are light trans-iron elements, not $r$-process elements as suggested in previous works.


A word of caution is needed; this study itself cannot answer satisfactorily to a question: ``how much $r$-process elements were made?". This is due to the fact that the choice of a reference abundance distribution was arbitrary and our conclusion strongly relied upon the abundances of only two radioactive isotopes $^{66}$Ni and $^{72}$Zn. The inferred mass of $r$-process elements ($\approx 0.01$; Table~2) merely reflects our choice of the reference abundance distribution, i.e., the $r$-process residuals to the solar system abundances. Comparison of our model (mFE-a) with the observed bolometric luminosity only ensures the production of light trans-iron isotopes $^{66}$Ni and $^{72}$Zn with the amounts presented in Table~\ref{tab:properties} (5th and 6th columns)\footnote{The light curve cannot disentangle the contributions of two decay chains from $^{66}$Ni and $^{72}$Zn because of their similar lifetimes. Measurements of $\gamma$-ray lines (97.8~keV and 145~keV/2.77~MeV, respectively) from a future nearby event may enable us to directly determine their abundances \citep[see, e.g.,][]{Hotokezaka2016}.} but the abundance distribution in Figure~\ref{fig:nuclei} (top-left)\footnote{For instance, an addition of ejecta with $Y_\mathrm{e} = 0.31$-0.35 (that were almost absent in our cases; Figure~\ref{fig:ye}, bottom) would overproduce the nuclei of $A \sim 100$ (Figure~\ref{fig:nuclei}, top) that had, however, little effect on radioactive heating (Figure~\ref{fig:qdot}, bottom). Similarly, additional ejecta with $Y_\mathrm{e} > 0.45$ might have little effect on heating (except for the decay chain from $^{56}$Ni at late times).}. Different from a previous thought, $\beta$-decay contribution from $A\sim 130$ is unimportant in this case (Figure~\ref{fig:qdot_isotope}, top left). Our result (mFE-a) is consistent with additional contributions from fission and $\alpha$-decay (Figure~\ref{fig:luminosity}, middle left) but only at a qualitative level when considering uncertainties in nuclear ingredients.

Currently, therefore, only the inferred $X_\mathrm{l}$ gives a hint to the amount of $r$-process elements. Our result in mFE-a, $X_\mathrm{l} = 0.035$, is a few times greater than the upper bound of the literature values  \citep[$\approx 10^{-4}$--$10^{-2}$, e.g.,][]{Arcavi2017, Chornock2017, Nicholl2017, Waxman2017}. If we took it literally, the $r$-process mass of $\approx 0.01\, M_\odot$ in mFE-a would be regarded as the upper limit. The true value may be a few times smaller, which is still consistent with the theoretical range of dynamical ejecta masses in a recent work \citep[$= 0.002$--0.016,][]{Shibata2017}. However, no information can be obtained from the inferred $X_\mathrm{l}$ alone on the prodution of heavy $r$-process elements such as gold and uranium. Only a viable strategy appears to search a signature of spontaneous fission (from $^{254}$Cf and possibly a few Fm isotopes), which can affect late-time luminosities \citep[][]{Wanajo2014, Hotokezaka2016, Zhu2018}. As described in \S~\ref{sec:heating},  many uncertainties are involved in estimating the fission contribution. Nevertheless, a variation of late-time light curves among future NS merger events will be indicative of actinide production (Figure~\ref{fig:luminosity}, grey area), which is expected from spectroscopic studies of Galactic halo stars.

In this paper comparison was made only with the NS merger GW170817, the first and currently unique detection of such an event. It should be noted that the same heating rates of our model (mFE-a) may not necessarily be applicable to future NS merger events. The reason is that the abundance distribution of light trans-elements may not be robust, which is very sensitive to $S$ and $Y_\mathrm{e}$ \citep{Meyer1998, Wanajo2018}. Small shifts of these quantities in post-merger disk outflows would substantially modify the abundance pattern. As can be seen in Figure~\ref{fig:qdot} (bottom), the behaviors of heating rates are very different among different $Y_\mathrm{e}$ groups for $Y_\mathrm{e} > 0.30$. The amounts of ejecta mass from disk outflows can also be dependent of, e.g., the masses and their ratio of merging NSs \citep[e.g.,][]{Shibata2017}. Even so, it is encouraging that we have means of discriminating between light trans-iron and $r$-process dominant mergers through the light curves of kilonovae in the future.

\acknowledgements

This work was supported by the JSPS
Grants-in-Aid for Scientific Research (26400232, 26400237), JSPS and CNRS under the Japan-France Research Cooperative Program, and the RIKEN iTHEMS Project.


\begin{thebibliography}{}
\bibitem[Abbott et al.(2017)]{Abbott+2017}
Abbott, B. P., Abbott, R., Abbott, T. D., et al. 2017, \prl, 119, 161101
\bibitem[Arcavi et al.(2017)]{Arcavi2017}
 Arcavi, I., Hosseinzadeh, G., Howell, D. A., et al. 2017, \nat, 551, 64
 \bibitem[Arnett(1979)]{Arnett1979}
 Arnett, W. D. 1979, \apj, 230, L37
\bibitem[Barnes \& Kasen(2013)]{Banes2013}
 Barnes, J., \& Kasen, D. 2013, \apj, 775, 18
\bibitem[Barnes et al.(2016)]{Barnes2016}
 Barnes, J., Kasen, D., Wu, M.-R., \& Mart\'inez-Pinedo, G. 2016, \apj, 829, 110 
\bibitem[Bauswein et al.(2013)]{Bauswein2013}
 Bauswein, A., Goriely, S., \& Janka, H.-T. 2013, \apj, 773, 78
\bibitem[Beniamini et al.(2016)]{Beniamini2016}
 Beniamini, P., Hotokezaka, K., \& Piran, T. 2016, \apjl, 829, L13
\bibitem[Bouquelle et al.(1996)]{Bouquelle1996}
 Bouquelle, V., Cerf, N., Arnould, M., Tachibana, T., \& Goriely S. 1996, \aap, 305, 1005
\bibitem[Burbidge et al.(1956)]{Burbidge1956}
 Burbidge, G. R., Hoyle, F., Burbidge, E. M., Christy, R. F., \& Fowler, W. A. 1956, Phys. Rev., 103, 1145
\bibitem[Burbidge et al.(1957)]{Burbidge1957}
 Burbidge, E. M., Burbidge, G. R., Fowler, W. A., \& Hoyle, F. 1957, Rev. Mod. Phys., 29, 547
\bibitem[Cameron(1957)]{Cameron1957}
 Cameron, A. G. W. 1957, Chalk River Report, CRL-41
\bibitem[Cardall \& Fuller(1997)]{Cardall1997}
 Cardall, C. Y., \& Fuller, G. M. 1997, \apjl, 486, L111
\bibitem[Chornock et al.(2017)]{Chornock2017}
 Chornock, R., Berger, E., Kasen, D., et al. 2017, \apjl, 848, L19
\bibitem[Colgate(1969)]{Colgate1969}
 Colgate, S. \& McKee, C. 1969, \apj, 157, 623 
\bibitem[C\^{o}t\'{e} et al.(2017)]{Cote2017}
C\^{o}t\'{e}, B., Belczynski, K., Fryer, C., et al. 2017, ApJ, 836, 230
\bibitem[Cowperthwaite et al.(2017)]{Cowperthwaite2017}
 Cowperthwaite, P. S., Berger, E., Villar, V. A., et al. 2017, \apjl, 848, L17
\bibitem[Cyburt et al.(2010)]{Cyburt2010}
  Cyburt, R. H., et al. 2010, \apjs, 189, 240
\bibitem[Dessart et al.(2009)]{Dessart2009}
 Dessart, L., Ott, C., Burrows, A., Rosswog, S., \& Livne, E. 2009, \apj, 690, 1681
\bibitem[Eichler et al.(1989)]{Eichler1989}
 Eichler, D., Livio, M., Piran, T., \& Schramm, D. N. 1989, Natur, 340, 126 
\bibitem[Farouqi et al.(2010)]{Farouqi2010}
 Farouqi, K., Kratz, K.-L., Pfeiffer, B., et al. 2010, \apj, 712, 1359
\bibitem[Fern\'andez et al.(2018)]{Fernandez2018}
 Fern\'andez, R., Tchekhovskoy, A., Quataert, E., Foucart, F., \& Kasen, D. 2018, \mnras, submitted	
\bibitem[Freiburghaus et al.(1999)]{Freiburghaus1999}
 Freiburghaus, C., Rosswog, S., \& Thielemann, F.-K. 1999, \apjl, 525, L121
\bibitem[Fujibayashi et al.(2018)]{Fujibayashi2018}
 Fujibayashi, S., Kiuchi, K., Nishimura, N., Sekiguchi, Y., \& Shibata, M. 2018, \apj, 860, 64	
\bibitem[Goriely \& Arnould(1996)]{Goriely1996}
 Goriely, S. \& Arnould, M. 1996, \aap, 312, 327
\bibitem[Goriely(1999)]{Goriely1999}
 Goriely, S. 1999, \aap, 342, 881
\bibitem[Goriely et al.(2007)]{Goriely2007}
 Goriely, S., Samyn, M., \& Pearson, J. M. 2007, \prc, 75, 064312
\bibitem[Goriely et al.(2008)]{Goriely2008}
 Goriely, S., Hilaire, S., \& Koning, A. J. 2008, \aap, 487, 767
\bibitem[Goriely et al.(2010)]{Goriely2010}
 Goriely, S., Chamel, N., \& Pearson, J. M. 2010, \prc, 82, 035804
\bibitem[Goriely et al.(2011)]{Goriely2011}
 Goriely, S., Bauswein, A., \& Janka, H.-T. 2011, \apjl, 738, L32
\bibitem[Goriely et al.(2015)]{Goriely2015}
 Goriely, S., Bauswein, A., Just, O., Pllumbi, E., \& Janka, H.-T. 2015, \mnras, 452, 3894
\bibitem[Goriely(2015)]{Goriely2015b}
 Goriely, S. 2015, Eur. Phys. J. A, 51, 22
\bibitem[Halevi \& M\"osta(2018)]{Halevi2018}
 Halevi, G. \& M\"osta, P. 2018, \mnras, 477, 2366
\bibitem[Hartmann et al.(1985)]{Hartmann1985}
 Hartmann, D., Woosley, S. E., \& El Eid, M. F. 1985, \apj, 297, 837
\bibitem[He{\ss}berger(2017)]{Hessberger2017}
 He{\ss}berger, F. P. 2017, Eur. Phys. J. A, 53, 75
\bibitem[Hillebrandt et al.(1976)]{Hillebrandt1976}
 Hillebrandt, W., Takahashi, K., \& Kodama, T. 1976, \aap, 52, 63
\bibitem[Holmbeck(2018)]{Holmbeck2018}
 Holmbeck, E. M., Beers, T. C., Roederer, I. U., et al. 2018, \apjl, 859, L24
\bibitem[Hotokezaka et al.(2013)]{Hotokezaka2013}
 Hotokezaka, K., Kiuchi, K., Kyutoku, K., et al. 2013, \prd, 87, 024001
\bibitem[Hotokezaka et al.(2016)]{Hotokezaka2016}
 Hotokezaka, K., Wanajo, S., Tanaka, M., et al. 2016, \mnras, 459, 35
\bibitem[Hotokezaka et al.(2017)]{Hotokezaka2017}
 Hotokezaka, K., Sari, R., \& Piran, T. 2017, \mnras, 468, 91
\bibitem[Ishimaru et al.(2015)]{Ishimaru2015}
 Ishimaru, Y., Wanajo, S., \& Prantzos, N. 2015, \apjl, 804, L35
\bibitem[Janka et al.(2012)]{Janka2012}
 Janka, H.-T., Hanke, F., Hüdepohl, L., et al. 2012, PTEP, 2012, 01A309
\bibitem[Ji et al.(2016)]{Ji2016}
 Ji, A., Frebel, A., Chiti, A., \& Simon, J. D. 2016a, \nat, 531, 610
\bibitem[Ji \& Frebel(2018)]{Ji2018}
 Ji, A. P. \& Frebel, A.2018, \apj, 856, 138
\bibitem[Just et al.(2015)]{Just2015}
 Just, O., Bauswein, A., Pulpillo, R. A., Goriely, S., \& Janka, H.-T., 2015, \mnras, 448, 541
\bibitem[Kasen et al.(2013)]{Kasen2013}
 Kasen, D., Badnell, N. R., \& Barnes, J. 2013, \apj, 774, 25
\bibitem[Kasen et al.(2015)]{Kasen2015}
 Kasen, D., Fernández, R., \& Metzger, B. D. 2015, \mnras, 450, 1777
\bibitem[Kasen et al.(2017)]{Kasen2017}
 Kasen, D., Metzger, B., Barnes, J., Quataert, E., \& Ramirez-Ruiz, E. 2017, \nat, 551, 80
\bibitem[Kasliwal et al.(2017)]{Kasliwal2017}
 Kasliwal, M. M., Nakar, E., Singer, L. P., et al. 2017, Sci., 358, 1559
\bibitem[Kitaura et al.(2006)]{Kitaura2006}
 Kitaura, F. S., Janka, H.-Th., \& Hillebrandt, W. 2006, \aap,
 450, 345
\bibitem[Korobkin et al.(2012)]{Korobkin2012}
 Korobkin, O., Rosswog, S., Arcones, A., \& Winteler, C. 2012, \mnras, 426, 1940
\bibitem[Lattimer \& Schramm(1974)]{Lattimer1974}
 Lattimer, J. M., \& Schramm, D. N. 1974, \apj, 192, L145
\bibitem[Li \& Paczy\'nski(1998)]{Li1998}
 Li, L.-X., \& Paczy\'nski, B. 1998, \apjl, 507, L59
\bibitem[Lippuner \& Roberts(2015)]{Lippuner2015}
 Lippuner, J., \& Roberts, L. F. 2015, \apj, 815, 82
\bibitem[Lippuner et al.(2017)]{Lippuner2017}
 Lippuner, J., Fern\'andez, R., Roberts, L. F., et al. 2017, \mnras, 472, 904
\bibitem[Lodders(2003)]{Lodders2003}
 Lodders, K. 2003, \apj, 591, 1220
\bibitem[Metzger et al.(2008)]{Metzger2008}
 Metzger, B. D., Piro, A. L., \& Quataert, E. 2008, \mnras, 390, 781
\bibitem[Metzger et al.(2010)]{Metzger2010}
 Metzger, B. D., Mart\'inez-Pinedo, G., Darbha, S., et al. 2010, \mnras, 406, 2650
\bibitem[Metzger \& Fern\'andez(2014)]{Metzger2014}
 Metzger, B. D., \& Fern\'andez, R. 2014, \mnras, 441, 3444
\bibitem[Metzger(2017)]{Metzger2017}
 Metzger, B. D. 2017, arXiv:1710.05931
\bibitem[Metzger(2018)]{Metzger2018}
 Metzger, B. D., Thompson, T. A., \& Quataert, E. 2018, \apj, 856, 101
\bibitem[Meyer(1989)]{Meyer1989}
 Meyer, B. S. 1989, \apj, 343, 254
\bibitem[Meyer et al.(1992)]{Meyer1992}
 Meyer, B. S., Mathews, G. J., Howard, W. M., Woosley, S. E., \& Hoffman, R. D. 1992, \apj, 399, 656
\bibitem[Meyer et al.(1996)]{Meyer1996}
 Meyer, B. S., Krishnan, T. D., \& Clayton, D. D. 1996, \apj, 462, 825
\bibitem[Meyer et al.(1998)]{Meyer1998}
  Meyer, B. S., Krishnan, T. D., \& Clayton, D. D. 1998, \apj, 498, 808
\bibitem[M\"osta et al.(2017)]{Moesta2017}
 M\"osta, P., Roberts, L. F., Halevi, G., et al. 2017, \apj, submitted; arXiv:1712.09370
\bibitem[Nicholl et al.(2017)]{Nicholl2017}
 Nicholl, M., Berger, E., Kasen, D., et al. 2017, \apjl, 848, L18
\bibitem[Nishimura et al.(2015)]{Nishimura2015}
 Nishimura, N., Takiwaki, T., \& Thielemann, F.-K. 2015, \apj, 810, 109
\bibitem[Nishimura et al.(2017)]{Nishimura2017}
 Nishimura N., Sawai H., Takiwaki T., Yamada S., \& Thielemann F.-K., 2017, \apjl, 836, L21
\bibitem[Ojima et al.(2018)]{Ojima2018}
 Ojima, T., Ishimaru, Y., Wanajo, S., Prantzos, N., \& Fran\c{c}ois, P. 2018, \apj, submitted
\bibitem[Otsuki et al.(2000)]{Otsuki2000}
 Otsuki, K., Tagoshi, H., Kajino, T., \& Wanajo, S. 2000, \apj, 533, 424
\bibitem[Papenfort et al.(2018)]{Papenfort2018}
 Papenfort, L. J., Gold, R., \& Rezzolla, L. 2018; arXiv1807.03795
\bibitem[Perego et al.(2015)]{Perego2015}
 Perego, A., Hempel, M., Fr\"olich, C., et al. 2015, \apj, 806, 275
\bibitem[Pian et al.(2017)]{Pian2017}
 Pian, E., D’Avanzo, P., Benetti, S., et al. 2017, \nat, 551, 67
\bibitem[Qian \& Woosley(1996)]{Qian1996}
 Qian, Y.-Z. \& Woosley, S. E. 1996, \apj, 471, 331
\bibitem[Rampp \& Janka(2002)]{Rampp2002}
 Rampp, M. \& Janka, H.-Th., 2002, \aap, 396, 361
\bibitem[Radice et al.(2016)]{Radice2016}
Radice, D., Galeazzi, F., Lippuner, J., et al. 2016, MNRAS, 460, 3255
\bibitem[Ruffert \& Janka(1999)]{Ruffert1999}
 Ruffert, M. \& Janka, H.-T. 1999, \aap, 344, 573
\bibitem[Safarzadeh \& Scannapieco(2017)]{Safarzadeh2017}
Safarzadeh, M., \& Scannapieco, E. 2017, \mnras, 471, 2088
\bibitem[Sato(1974)]{Sato1974}
 Sato, K. 1974, Prog. Theor. Phys., 51, 726
\bibitem[Schramm(1973)]{Schramm1973}
 Schramm, D. N. 1973, \apj, 185, 293 434, 102
\bibitem[Sekiguchi et al.(2015)]{Sekiguchi2015}
 Sekiguchi, Y., Kiuchi, K., Kyutoku, K., \& Shibata, M. 2015, \prd, 91, 064059
\bibitem[Sekiguchi et al.(2016)]{Sekiguchi2016}
 Sekiguchi, Y., Kiuchi, K., Kyutoku, K., Shibata, M., \& Taniguchi, K. 2016, \prd, 93, 124046
\bibitem[Siegel \& Metzger(2017)]{Siegel2017}
 Siegel, D. \& Metzger, B. D. 2017, \prl, 119, 1102
\bibitem[Shibata et al.(2017)]{Shibata2017}
 Shibata, M., Fujibayashi, S., Hotokezaka, K., et al. 2017, \prd, 96, 123012
\bibitem[Siqueira Mello et al.(2013)]{Siqueira2013}
 Siqueira Mello, C., Jr., Spite, M., Barbuy, B., et al. 2013, \aap, 550, A122
\bibitem[Smartt et al.(2017)]{Smartt2017}
 Smartt, S. J., Chen, T.-W., Jerkstrand, A., et al. 2017, \nat, 551, 75
\bibitem[Sneden et al.(2003)]{Sneden2003}
 Sneden, C., Cowan, J. J., Lawler, J. E., et al. 2003, \apj, 591, 936
\bibitem[Sneden et al.(2008)]{Sneden2008}
 Sneden, C., Cowan, J. J., \& Gallino, R. 2008, \araa, 46, 241
\bibitem[Symbalisty \& Schramm(1982)]{Symbalisty1982}
 Symbalisty, E., \& Schramm, D. N. 1982, ApL, 22, 143
\bibitem[Sumiyoshi et al.(2001)]{Sumiyoshi2001}
 Sumiyoshi, K., Terasawa, M., Mathews, G. J., et al. 2001, \apj, 562, 880
\bibitem[Surman et al.(2008)]{Surman2008}
 Surman, R., McLaughlin, G. C., Ruffert, M., Janka, H.-Th., \& Hix, W. R. 2008, \apjl, 679, L117
\bibitem[Tachibana et al.(1990)]{Tachibana1990}
 Tachibana, T., Yamada, M.,\& Yoshida, Y. 1990, PThPh, 84, 641
\bibitem[Tanaka \& Hotokezaka(2013)]{Tanaka2013}
 Tanaka, M., \& Hotokezaka, K. 2013, \apj, 775, 113
\bibitem[Tanaka et al.(2017)]{Tanaka2017}
 Tanaka, M., et al. 2017, Publ. Astron. Soc. Jap., arXiv:1710.05850
\bibitem[Tanaka et al.(2018)]{Tanaka2018}
 Tanaka, M., Kato, D., Gaigalas, G., et al. 2018, \apj, 852, 109
\bibitem[Tanvir et al.(2017)]{Tanvir2017}
 Tanvir, N. R., Levan, A. J., Gonz\'alez-Fern\'andez, C., et al. 2017, \apjl, 848, L27
\bibitem[Timmes \& Swesty(2000)]{Timmes2000}
 Timmes, F. X., \& Swesty, F. D. 2000, \apjs, 126, 501
\bibitem[Thielemann et al.(2017)]{Thielemann2017}
 Thielemann, F.-K., Eichler, M., Panov, I. V., \& Wehmeyer, B. 2017, ARNPS, 67, 253
\bibitem[Thompson(2001)]{Thompson2001}
 Thompson, T. A., Burrows, A., \& Meyer, B. S. 2001, \apj, 562, 887
\bibitem[Villar et al.(2018)]{Villar2018}
 Villar, V. A., Cowperthwaite, P. S., Berger, E., et al. 2018, \apjl, 862, L11
\bibitem[Viola et al.(1985)]{Viola1985}
 Viola, V. E., Kwiatkowski, K. \& Walker, M. 1985, \prc, 31, 1550
\bibitem[Wanajo et al.(2001)]{Wanajo2001}
 Wanajo, S., Kajino, T., Mathews, G. J., \& Otsuki, K. 2001, \apj, 554, 578
\bibitem[Wanajo et al.(2003)]{Wanajo2003}
 Wanajo, S., Tamamura, M., Itoh, N., et al. 2003, \apj, 593, 968
\bibitem[Wanajo et al.(2011)]{Wanajo2011}
 Wanajo, S., Janka, H.-T., \& M\"uller, B. 2011, \apj, 726, L15
\bibitem[Wanajo \& Janka(2012)]{Wanajo2012}
 Wanajo, S., \& Janka, H.-T. 2012, \apj, 746, 180
\bibitem[Wanajo et al.(2013)]{Wanajo2013}
 Wanajo, S., Janka, H.-T., \& M\"uller, B. 2013, ApJL, 767, L26
\bibitem[Wanajo(2013)]{Wanajo2013b}
 Wanajo, S. 2013, ApJL, 770, L22
\bibitem[Wanajo et al.(2014)]{Wanajo2014}
 Wanajo, S., Sekiguchi, Y., Nishimura, N., et al. 2014, \apjl, 789, L39
\bibitem[Wanajo et al.(2018)]{Wanajo2018}
 Wanajo, S., M\"uller, B., Janka, H.-T., \& Heger, A. 2018, \apj, 852, 40
\bibitem[Waxman et al.(2017)]{Waxman2017}
 Waxman, E., Ofek, E., Kushnir, D., \& Gal-Yam, A. 2017, arXiv:1711.09638
\bibitem[Winteler et al.(2012)]{Winteler2012}
 Winteler, C., et al. 2012, \apjl, 750, L22
\bibitem[Witti et al.(1994)]{Witti1994}
 Witti, J., Janka, H.-Th., \& Takahashi, K. 1994, \aap, 286, 841
\bibitem[Woosley \& Hoffman(1992)]{Woosley1992}
 Woosley, S. E. \& Hoffman, R. D. 1992, \apj, 395, 202
\bibitem[Woosley et al.(1994)]{Woosley1994}
 Woosley, S. E., Wilson, J. R., Mathews, G. J., Hoffman, R. D., \&
 Meyer, B. S.  1994, \apj, 433, 229
\bibitem[Woosley(1997)]{Woosley1997}
 Woosley, S. E. 1997, \apj, 476, 801
\bibitem[Wu et al.(2016)]{Wu2016}
Wu, M.-R., Fern\'andez, R., Mart\'inez-Pinedo, G., \& Metzger, B. D. 2016, \mnras, 463, 2323
\bibitem[Zhu et al.(2018)]{Zhu2018}
 Zhu, Y., Wollaeger, R. T., Vassh, N., et al. 2018, arXiv:1806.09724
\end{thebibliography}
\end{document}